\documentclass[iop,usenatbib]{emulateapj}

\usepackage{epsfig}

\def\HI{\hbox{H\hskip1.5pt$\scriptstyle\rm I\ $}}
\def\HIp{\hbox{H\hskip1.5pt$\scriptstyle\rm I$}}
\def\HII{\hbox{H\hskip1.5pt$\scriptstyle\rm II\ $}}
\def\HIIp{\hbox{H\hskip1.5pt$\scriptstyle\rm II$}}
\def\HeI{\hbox{He\hskip1.5pt$\scriptstyle\rm I\ $}}
\def\HeIp{\hbox{He\hskip1.5pt$\scriptstyle\rm I$}}
\def\HeII{\hbox{He\hskip1.5pt$\scriptstyle{\rm II}\ $}}
\def\HeIIp{\hbox{He\hskip1.5pt$\scriptstyle{\rm II}$}}

\def\HeIII{\hbox{He\hskip1.5pt$\scriptstyle\rm III\ $}}
\def\HeIIIp{\hbox{He\hskip1.5pt$\scriptstyle\rm III$}}

\def\nHII{_{\rm HII}}

\def\nHeIII{_{\rm HeIII}}

\def\C{_{\rm c}}
\def\J{_{\rm J}}

\def\III{_{\rm III}}
\def\Si{_{\rm SII}}
\def\bSi{_{\rm \!\!SII}}
\def\se{^{\rm e}}
\def\H{_{\rm H}}

\def\HH{H$_2$\ }

\def\HHp{H$_2$}
\def\nHH{_{\rm H_2}}

\def\jj{_{\rm j}}

\def\J{_{\rm J}}

\def\DB{_{\rm DB}}

\def\dis{_{\rm dis}}

\def\lya{Ly$\alpha$}

\def\re-ion{_{\rm ion}}
\def\rec{_{\rm rec}}

\def\der{{\rm d}}
\def\sec{Sec.\ }

\def\T{_{\rm T}}

\def\b{_{\rm b}}

\def\g{_{\rm g}}

\def\emi{_{\rm em}}
\def\maxi{_{\rm max}}

\def\modot{M$_\odot$~}

\def\modotd{M$_\odot$}

\def\lav{\langle}
\def\rav{\rangle}

\def\maxi{_{\rm max}}

\def\rec{_{\rm rec}}

\def\s{_{\rm s}}

\def\hg{_{\rm hg}}
\def\rh{_{\rm rh}}  
\def\rhd{_{\rm rh,D}}  
\def\rhb{_{\rm rh,B}}  
\def\rhc{_{\rm rh,C}}

\def\cg{_{\rm cg}}

\def\cgb{_{\rm cg,B}}
\def\cgc{_{\rm cg,C}}

\def\up{_{\rm up}}

\def\SN{_{\rm SN}}

\def\AGN{_{\rm AGN}}

\def\sff{_{\rm sf}}
\def\sfd{_{\rm sf,D}}
\def\sb{_{\rm s,B}}
\def\sfb{_{\rm sf,B}}
\def\sc{_{\rm s,C}}
\def\sfc{_{\rm sf,C}}
\def\cool{_{\rm cool}}
\def\coold{_{\rm cool,D}}
\def\coolb{_{\rm cool,B}}
\def\coolc{_{\rm cool,C}}

\def\dynb{_{\rm dyn,B}}
\def\dync{_{\rm dyn,C}}

\def\IRA{_{\rm eff}}
\def\IGM{_{\rm IGM}}
\def\crit{_{\rm crit}}

\def\obs{_{\rm obs}}
\def\acc{_{\rm acc}}

\def\lossc{_{\rm sl,X}}

\def\eff{_{\rm eff}}
\def\PhiIRA{C\eff}

\def\emi{_{\rm em}}
\def\esc{_{\rm esc}}
\def\ch{_{\rm c}}

\def\lim{_{\rm lim}}

\def\IGM{_{\rm IGM}}

\def\c{_{\rm C}}
\def\d{_{\rm D}}

\def\g{_{\rm hg}}
\def\si{_{\rm i}}

\def\p{_{\rm p}}
\def\s{_{\rm s}}

\def\B{_{\rm B}}
\def\C{_{\rm C}}
\def\D{_{\rm D}}

\def\G{_{\rm G}}
\def\H{_{\rm H}}

\def\J{_{\rm J}}

\def\ui{^{(i)}}
\def\u1{^{(1)}}
\def\un1{^{(i\ne 1)}}
\def\gi{^{\rm g~(i)}}
\def\si{^{\rm s~(i)}}
\def\in{^{\rm in}}
\def\out{^{\rm out}}

\def\lya{Ly$\alpha$}

\def\HH{H$_2$\ }

\def\HI{\hbox{H\hskip1.pt$\scriptstyle\rm I$\ }}

\def\HII{\hbox{H\hskip1.pt$\scriptstyle\rm II$\ }}

\def\HeI{\hbox{He\hskip1.pt$\scriptstyle\rm I$\ }}
\def\HeII{\hbox{He\hskip1.pt$\scriptstyle{\rm II}$\ }}
\def\nHH{_{\rm H_2}}
\def\nHII{_{\rm HII}}

\def\pIII{Pop III\ }

\def\pIaII{Pop I \& II\ }

\def\pIaIIf{Pop I \& II}

\def\diss{_{\rm diss}}
\def\shi{_{\rm sh}}
\def\kb{k_{\rm B}}
\def\y3{$p_{\rm III}$}
\def\yb3{$p_{\rm III}$\ }
\def\e3{$\epsilon_{\rm III}$}
\def\eb3{$\epsilon_{\rm III}$\ }

\def\rec{_{\rm rec}}

\def\der{{\rm d}}

\def\SN{_{\rm SN}}

\def\nbody{{$N$}-body\ }

\def\Omb{{\Omega_{\rm b}}}

\def\a{_{\rm a}}

\def\deli{\Delta_{\rm i}}

\def\lsim{\lower.5ex\hbox{\ltsima}}
\def\gsim{\lower.5ex\hbox{\gtsima}}
\def\la{\lsim}
\def\ga{\gsim}
\def\gtsima{$\; \buildrel > \over \sim \;$}
\def\ltsima{$\; \buildrel < \over \sim \;$}

\newcommand{\beq}{\begin{equation}}
\newcommand{\eeq}{\end{equation}}
\newcommand{\beqa}{\begin{eqnarray}}
\newcommand{\eeqa}{\end{eqnarray}}

\def\nbody{{$N$}-body }

\def\last{_{\rm last}}
\def\p{_{\rm p}}
\def\e{_{\rm e}}

\def\h{_{\rm h}}
\def\H{_{\rm H}}

\def\c{_{\rm c}}

\def\hg{_{\rm hg}}
\def\rh{_{\rm rh}}  
\def\rhd{_{\rm rh,D}}  
\def\rhb{_{\rm rh,B}}  
\def\rhc{_{\rm rh,C}}  
  
\def\cg{_{\rm cg}}

\def\cgb{_{\rm cg,B}}
\def\cgc{_{\rm cg,C}}

\def\up{_{\rm up}}
\def\SN{_{\rm SN}}

\def\sfd{_{\rm sf,D}}
\def\sb{_{\rm s,B}}
\def\sfb{_{\rm sf,B}}
\def\sc{_{\rm s,C}}
\def\sfc{_{\rm sf,C}}
\def\cool{_{\rm cool}}
\def\coold{_{\rm cool,D}}
\def\coolb{_{\rm cool,B}}
\def\coolc{_{\rm cool,C}}

\def\dynb{_{\rm dyn,B}}
\def\dync{_{\rm dyn,C}}
\def\IRA{_{\rm eff}}
\def\BH{_{\rm BH}}
\def\BHc{_{\rm BH,C}}
\def\BHb{_{\rm BH,B}}
\def\BHd{_{\rm BH,D}}

\def\crit{_{\rm crit}}

\def\obs{_{\rm obs}}
\def\acc{_{\rm acc}}

\def\lossc{_{\rm loss,C}}

\def\eff{_{\rm eff}}
\def\PhiIRA{C\eff}

\begin{document}

\shorttitle{AMIGA}
\shortauthors{Manrique et al.}

\title{Leaving the dark ages with AMIGA}

\author{Alberto Manrique$^{1}$\footnote{E-mail: a.manrique@ub.edu},
  Eduard Salvador-Sol\'e$^{1}$, Enric Juan$^{1}$, Evanthia
  Hatziminaoglou$^{2}$, Jos\'e Mar\'\i a Rozas$^{1}$, Antoni
  Sagrist\`a$^{1}$, Kevin Casteels$^{1}$, Gustavo Bruzual$^{3}$,
  and Gladis Magris$^{4}$} \affil{$^{1}$Institut de Ci\`encies del
  Cosmos. Universitat de Barcelona, UB-IEEC. Mart\'\i\ i Franqu\`es 1,
  E-08028 Barcelona, Spain} \affil{$^{2}$European Southern
  Observatory, Karl-Schwarzschild-Str. 2, 85748 Garching bei München,
  Germany} \affil{$^{3}$Centro de Radioastronom\'\i a y Astrof\'\i
  sica, UNAM, Campus Morella, M\'exico} \affil{$^{4}$Centro de
  Investigaciones de Astronom\'\i a, Apartado Postal 264, M\'erida
  5101-A, Venezuela}

\begin{abstract}

We present an {\it Analytic Model of Intergalactic-medium and GAlaxy}
evolution since the dark ages. AMIGA is in the spirit of the popular
semi-analytic models of galaxy formation, although it does not use
halo merger trees but interpolates halo properties in grids that are
progressively built. This strategy is less memory-demanding and allows
one to start the modeling at redshifts high enough and halo masses low
enough to have trivial boundary conditions. The number of free
parameters is minimized by making the causal connection between
physical processes usually treated as independent from each other,
which leads to more reliable predictions. But the strongest points of
AMIGA are: i) the inclusion of molecular cooling and metal-poor,
population III (Pop III) stars, with the most dramatic feedback, and
ii) the accurate follow-up of the temperature and volume filling
factor of neutral, singly, and doubly ionized regions, taking into
account the distinct halo mass functions in those environments.  We
find the following general results. Massive Pop III stars determine
the IGM metallicity and temperature, and the growth of spheroids and
disks is self-regulated by that of massive black holes developed from
the remnants of those stars. Yet, the properties of normal galaxies
and active galactic nuclei appear to be quite insensitive to Pop III
star properties owing to the much higher yield of ordinary stars
compared to Pop III stars and the dramatic growth of MBHs when normal
galaxies begin to develop, which cause the memory loss of the initial
conditions.

\end{abstract}

\keywords{galaxies --- galaxies: formation --- dark matter ---
  intergalactic medium}

\section{INTRODUCTION}\label{int}

Galaxies develop within dark matter (DM) halos through mergers and gas
accretion. This ``hierarchical scenario''
\citep{RO77,S77,WR78,Betal84,WF91} explains indeed the main observed
galaxy properties. However, some aspects of the nearby universe resist 
being satisfactorily recovered (e.g. \citealt{Benson10,Cea06}), and
the increasing amount of data at progressively higher redshifts, $z$'s,
is permanently challenging our ideas within this theoretical
framework.

A great progress has been achieved in the last decades in this field
thanks to the use of hydrodynamic simulations
(e.g. \citealt{TLA97,SN99,Spring00,Nea04,Spring05,Sch10}) and
semi-analytic models (SAMs)
\citep{KWG,CAFNZ94,SP99,KCDW99,CLBF00,Hattea03,Bea03,Mea05,Bea06,Mea07,RF10,
  Fontea11}. SAMs are more flexible and inform more easily on the main
properties of objects. However, they have the reputation of describing
the baryon physics by means of too simple recipes and including too
many parameters. Simulations certainly provide more detailed
information and are, in principle, based on first principles. However,
they involve the same recipes and parameters as SAMs at
subresolution scales.

But all these tools suffer for the same limitations: the huge amounts
of memory and CPU time involved. This is annoying for two
reasons. Firstly, galaxy formation is a non-linear process where the
feedback of luminous objects on the intergalactic medium (IGM) plays a
central role (see e.g. \citealt{MSS14}, hereafter MSS, and references
therein). Yet, those limitations prevent from treating
self-consistently the coupling of these two baryonic phases. In
particular, the ionizing background, with important consequences for
dwarf galaxies (e.g. \citealt{Bea02,HOJN11,MEtal11}) must be treated
in an adhoc fashion. Secondly, galaxy properties depend on those of
their earlier low-mass progenitors. However, the highest redshift $z$
and the minimum halo mass $M\H$ that can be reached in studies of
nearby galaxies are about 7 and $10^9$ \modotd, respectively, both in
SAMs (e.g. \citealt{Bea06}) and simulations
(e.g. \citealt{Sch10}).\footnote{Even though current simulations start
  at $z\ga 100$, convergence of galaxy properties is only found up to
  $z\sim 7$ for the most favorable case of relatively low resolutions
  \citep{Sch10}.}  Of course, when studies focus either on small
regions or high-$z$'s, the limits are less stringent, although yet too
restrictive.

More importantly, the first generation stars formed by molecular
cooling from the original pristine gas, the so-called Population III
(Pop III) stars, are responsible for the initial metal enrichment and
reionization of the IGM, as well as for the seeds of massive black
holes (MBHs). Their local effects can be studied in detail by means of
high-resolution hydrodynamic simulations
\citep{WA07,WA08,Tea09,Stea10,Kea11,Pea11}. But the limited dynamic
range of simulations prevents from analyzing at the same time their
cosmological effects. The only attempt to date to account for the
feedback of \pIII stars in hydrodynamic simulations is due to
\citet{CEtal00}, while, in the case of SAMs better adapted in
principle to the study of galaxy formation on cosmological scales,
there is only the work by \citet{CF05}. Unfortunately, in both
studies, the baryon physics is dealt with by means of too simple
analytic recipes and galaxies are not realistically modeled.

Last but not least, IGM is a composite (several chemical species),
multiphase (singly and doubly ionized bubbles and subbubbles embedded
in a neutral background), inhomogeneous (density and temperature
fluctuations) environment, whose accurate analytic modeling is hard to
achieve without important simplifying assumptions. 

An improved analytic treatment of IGM has been recently developed by
MSS. In the present Paper, we couple it to AMIGA, an {\it Analytic
  Model of Intergalactic-medium and GAlaxy} evolution specifically
devised to monitor those cosmic components since the dark ages. AMIGA
includes molecular cooling and \pIII stars with the most dramatic
feedback. To save memory AMIGA does not rely on the construction of
individual halo merger trees but on the interpolation in grids of halo
properties that are progressively built, starting from well-known
boundary conditions. In addition, it makes the causal connection of
physical processes usually dealt with independently from each
other. This reduces the number of free parameters and leads to a model
which is internally more consistent than previous SAMs. Its
application to the study of reionization is given elsewhere
\citep{SM13}. Here, we describe the general model, putting special
emphasis in its novelties.

The outline of the paper is as follows. In Section \ref{strat}, we
describe the general procedure followed in AMIGA. Sections \ref{DM},
\ref{IGM}, \ref{stars}, \ref{galaxies}, and \ref{MBH} are respectively
devoted to the modeling of DM, gas, stars, galaxies and MBHs. In
Section \ref{mass}, we summarize the mass and metallicity evolution of
the different baryonic phases and galactic components, and, in Section
\ref{out}, we describe how the final photometric properties of
luminous objects are computed. Lastly, in Section \ref{summ} we
discuss the main achievements and some fundamental results of AMIGA.

The specific results shown throughout the paper correspond to
plausible values of the AMIGA parameters in the concordant
$\Lambda$CDM cosmology characterized by $\Omega_\Lambda=0.712$,
$\Omega_{\rm m}=0.288$, $\Omega_{\rm b}=0.0472$, $H_0 = 69.3$ km
s$^{-1}$ Mpc$^{-1}$, $n_{\rm s}=0.97$, and $\sigma_8 = 0.830$
\citep{HinEtal13}. Whenever possible, they are compared to
observational data in order to assess the goodness of the models. The
reader is referred to \citet{SM13} for detailed information on the
source of these data and the parameter values.


\section{GENERAL PROCEDURE}\label{strat}

To minimize the memory and CPU time requirements of AMIGA special
attention is paid in treating every random process entering the
problem in the best suited way. If it is such that every single event
has a noticeable, possibly critical effect, the random process is
dealt with in a full probabilistic fashion. Otherwise, it is dealt
with in a deterministic fashion, by calculating its secular action in
the desired time interval. In both cases, we use either analytic or
well-sampled numerical probability distribution functions (PDFs).

\begin{figure}
\vspace{0.25cm}
\centerline{\psfig{file=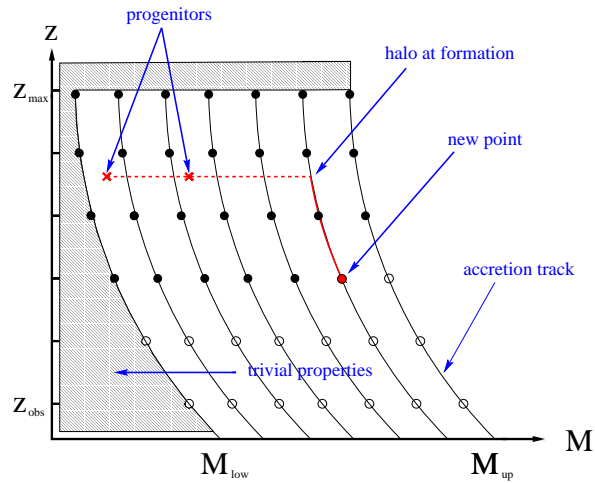,width=.35\textwidth,angle=-90}}
\vspace{0.2cm}
\caption{Cartoon representing how the interpolation grids of halo
  properties used in AMIGA are built. For simplicity we assume here
  that all progenitors are located in a given (neutral or ionized)
  environment and have identical age, so their properties can be found
  by interpolation in the piece of one only grid where all halos have
  that age. The shaded area represents the region where halos have
  trivial properties because they are not able to trap baryons, while
  halos with DM masses above $M_{\rm up}$ are highly improbable.}
\label{grid}
\end{figure}

AMIGA does not proceed by explicitly constructing Monte Carlo or
\nbody halo merger trees, but by interpolating the typical properties
of halos in neutral and ionized regions in two parallel 3D grids with
$n_{\rm z}$ log-bins of $1+z$, $n\H$ log-bins of DM masses $M\H$, and
$n_{\rm a}$ linear bins of halo ages $t\a\equiv t(z)-t\H$, where
$t(z)$ is the cosmic time corresponding to $z$ and $t\H$ is the halo
formation time, defined as the time of the last major merger having
caused the rearrangement of the system. These grids are progressively
built (see Fig.~\ref{grid}), starting at a redshift $z_{\rm max}$ high
enough for halos of all masses to have trivial properties down to the
redshift $z_{\rm obs}$ of observation; at every $z$, from a value of
$M\H$ low enough ($10$ \modot at $z_{\rm max}=60$) for halos to also
have trivial properties up to a value high enough ($10^{15}$ \modot at
$z_{\rm obs}=0$) for them to be highly improbable; and at every couple
of $z$ and $M\h$ values, for a set of halo ages spanning over the
relevant time interval. In this way, integrating at every $z$ over
halo ages for the halo formation time PDF, and over halo masses for
the appropriate halo mass function (MF), we determine the
instantaneous change induced by luminous sources in the IGM properties
at that $z$. This is a notable improvement compared to ordinary SAMs
where the feedback of luminous sources at a given $z$ is only known
for the small number of halo masses and ages covered by the discrete
branches of the merger tree that is being built.

\begin{figure*}
\centering
\vspace{-0.1cm}
\centerline{\psfig{file=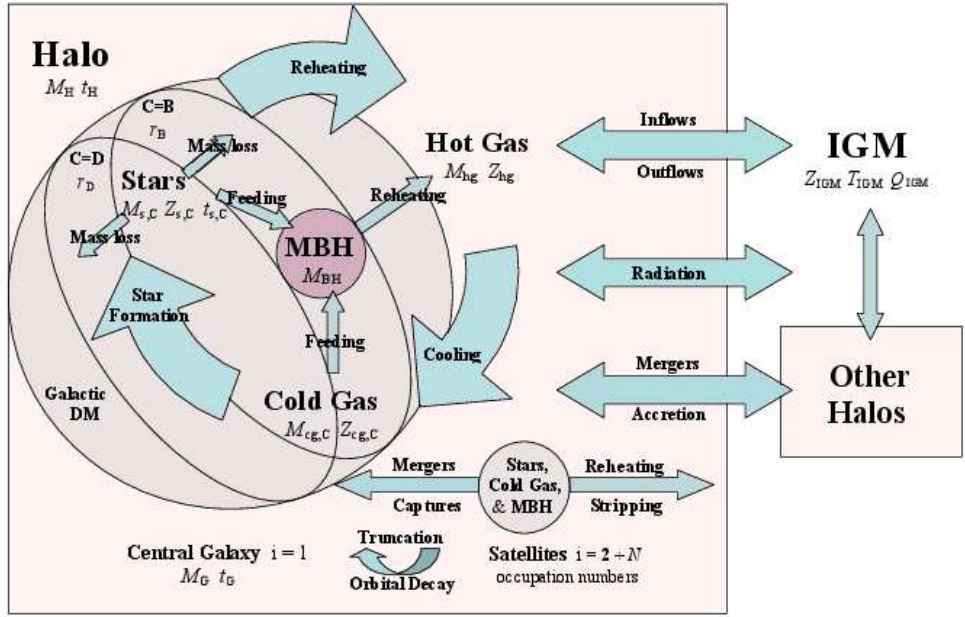,width=0.90\textwidth}}
\vspace{-1.cm}
\caption{Halo and IGM properties followed in AMIGA, and the physical
  processes involving them. The properties of halos at $z$, with DM
  mass $M\H$ and formation time $t\H$, stored in the interpolation
  grids for neutral and ionized regions are: the mass $M\hg$ and
  metallicity $Z\hg$ of the hot intrahalo gas, hereafter the hot gas,
  and the properties of the central galaxy (i $=1$), namely its total
  mass (including the galactic DM) $M\G$ and formation time $t\G$, the
  mass $M\cgc$ and metallicity $Z\cgc$ of the cold interstellar gas,
  hereafter the cold gas, the mass $M\sc$ and metallicity $Z\sc$ of
  stars formed at different times $t\sc$ (when these quantities refer
  to stars at formation, subscript s is replaced by sf) in the disk
  (C=D) and spheroid (C=B), their respective scale radii, $r\D$ and
  $r\B$, and the mass $M\BH$ of the central MBH. The properties of
  satellites (i $=2\div N$) are stored in the form of occupation numbers in
  a multidimensional space of galactic properties (essentially the
  same as for the central object). The metallicity $Z\IGM$ and
  temperature $T\IGM$ of the neutral, singly, and doubly ionized IGM
  phases, with respective volume filling factors $Q\IGM$, do not need
  to be stored in interpolation grids.}
\label{Defs}
\end{figure*}

To obtain the typical properties of a halo at a new point $(z,M\H,a)$
of a grid we chose the masses and formation times of its progenitors
according to the corresponding PDFs, and find their typical properties
by interpolation in the pieces of grids previously built covering all
possible halo ages, all halo masses less than $M\H$, and all redshifts
higher than $z$. From the properties of the progenitors, we determine
those of the halo at formation, and then follow its evolution by
continuous accretion until reaching the $z$ of the final object (see
Fig.~\ref{grid}). The accretion rate is given by the analytic halo
growth model (Sec.~\ref{DM}), and the composition at any moment of
accreted matter is well-known: (i) substantial halos whose properties
are obtained by interpolation within the grids, (ii) tiny halos with
trivial properties lying outside the grids, and (iii) a
well-determined fraction of non-trapped intergalactic gas (\sec
\ref{nt}). Finally, averaging the properties of the halo obtained from
each progenitor configuration, we obtain the quantities to be stored
in the new point of the grid.

All halo properties, including the baryonic content, stored at every
point of the grids and the main physical processes where they are
involved are represented, in the notation used throughout the paper,
in Figure \ref{Defs}. Taking advantage of the fact that satellites are
numerous, and hence, can be dealt with statistically, we do not store
the values of their individual properties as for the central galaxy,
but their occupation numbers in a multidimensional space of galactic
properties, with $n_{\rm f}$ linear bins of formation time, i.e. the
last moment the satellite structure was rearranged, $n_{\rm m}$
log-bins in total mass, $n_{\rm bm}$ log-bins in baryonic to total
mass ratio, $2\, n_{\rm sb}$ log-bins in disk and spheroid stellar to
total baryonic mass ratio, $3\, n_{\rm Z}$ log-bins in disk and
spheroid stellar metallicity, and disk gas metallicity, and $n_{\rm
  \sigma}$ log-bins in disk central surface density. This latter
property is used to calculate the disk scale radius given its mass,
while the spheroid scale radius is calculated making use of the
average dissipative contraction factor (see Sec.~\ref{mom}) of central
spheroids with identical stellar masses and formation times. Lastly,
the mass of satellite MBHs is calculated making use of the constant
average MBH to stellar mass ratio of central spheroids with identical
masses.

AMIGA is implemented in an OMP (shared memory) parallelized code with
32 CPUs. The time spent by a run depends mostly on the size of the
satellite array and the value of $z_{\rm obs}$ as the non-null
occupation numbers filling that array increase with decreasing
$z$. For $z_{\rm obs} =2$ and the minimum sizes of the interpolation
grids ($n_{\rm z}=51$, $n\H=91$, and $n_{\rm a}=3$) and of the
satellite array ($n_{\rm f}=8$, $n_{\rm m}=38$, $n_{\rm bm}=24$,
$n_{\rm sb}=6$, $n_{\rm Z}=4$, and $n_{\rm \sigma}=4$) ensuring
convergence, it takes about 76 hours. For such standard dimensions, the
typical properties of a halo in the grid arise from $3^2$ distinct
progenitor configurations, and the typical properties of galaxies in
halos with $M\H$ at $z$ arise from to $2\times 3^3$ different halo
progenitor configurations, the additional factor three arising from
the different halo ages, and the factor two from the fact that the
host halo may form either in a neutral region (before it harbors
galaxies) or in a ionized one.

\section{DARK MATTER}\label{DM}

At the time of matter-radiation equality, DM begins to cluster in
halos that merge with each other and grow from small to large
scales. Halos will become the backbone of galaxies, so it is mandatory
to accurately model their mass growth, inner structure, and
kinematics. This is achieved in AMIGA within the framework of the
excursion set formalism as in usual non-hybrid SAMs, but in a slightly
modified version of it, called the modified extended Press-Schechter
formalism (MEPS) \citep{ssm98,RGS01}, with important advantages
compared to the usual extended Press-Schechter (EPS) model.

\begin{figure}
\vspace{0.7cm}
\centerline{\psfig{file=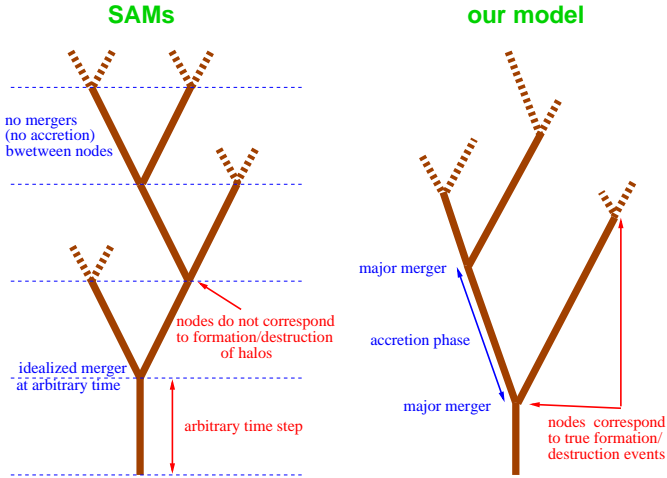,width=.35\textwidth,angle=-90}}
\vspace{0.75cm}
\caption{Comparison between the idealized halo merger trees, with
  adhoc finite resolution in both mass and time, implemented in
  ordinary non-hybrid SAMs, and the more realistic merger trees
  resulting from the MEPS formalism used in AMIGA.}
\label{trees}
\end{figure}

The conditional MF in the EPS model diverges in the limit of small
$M\H$. Thus, merger trees are infinitely ramified, which forces one to
adopt a finite resolution in mass and time. The finite resolution in
mass prevents from properly dealing with the capture of low-mass halos
contributing to accretion, and introduces some uncertainty in the
total number and mass of resolved progenitors \citep{SP99}. While the
finite resolution in time prevents from accurately dealing with
mergers because the conditional MF ensures only that a halo with $M\H$
at $t$ is found at $t' > t$ within another halo with $M\H'> M\H$; it
does not inform on the exact moment when the incorporation takes
place. Yet, the properties of a halo at a node of the merger tree are
inferred from those of its progenitors at the previous node, {\em
  assuming the merger takes place at that moment}, evolved until the
time of the final node. In other words, the timing and properties of
the evolving objects do not match those of the real merging process
(see Fig. \ref{trees}). To minimize the effects of such an inaccuracy
a relatively small time step must be adopted, but then the need of
storing all the information on the merger tree prevents from reaching
a very high-$z$.

The MEPS formalism making the distinction between minor and major
mergers does not have any of these drawbacks. Major mergers are really
binary \citep{RGS01}, so they can be dealt with statistically, in a
fully accurate fashion, with no need to introduce any limited
resolution. While minor mergers can also be dealt with fully
accurately through their global secular action. Specifically, one can
calculate the halo DM mass accretion rate, $\dot M\H$, as a function
of $M\H$ and $t$. In addition, the MEPS formalism allows one to
calculate the PDFs of halo formation times and progenitor masses
\citep{RGS01}, not available in the usual excursion set formalism. All
these differences lead to exact merger trees, with realistic discrete
branching (see Fig.~\ref{trees}). Furthermore, the MEPS formalism also
allows one to accurately derive the inner structure and kinematics of
virialized DM halos \citep{Sea07,Sea12a,SS12b}. We stress that,
contrarily to ordinary SAMs, all these halo properties are determined
in AMIGA directly from the cosmology considered with no single free
parameter.

AMIGA monitors the evolution of halos lying in neutral and ionized
regions, separately. Such a distinction is important because pristine
gas only falls inside halos lying in neutral regions; in ionized
regions, the IGM is polluted with metals produced in galaxies. The
feedback of luminous sources on the ionized IGM is computed using the
halo mass function in those ionized regions, which is slightly
different from that of halos in neutral regions (see MSS).

\section{The Gas}\label{IGM}

Until recombination at $z\sim 1100$, radiation pressure prevents the
ionized gas from falling into the halo potential wells. Nonetheless,
until $z=150(\Omb h^2/0.023)^{2/5}-1\simeq 150$, the abundance of free
electrons is high enough for the neutral gas to be kept thermalized
with the cosmic background radiation (CMB). At that $z$, the residual
abundance of free electrons ($\bar x_{\rm e}\approx 3.1\times
10^{-4}$) freezes out, and the gas begins to undergo adiabatic
cooling. At the beginning, the gas is too hot to be trapped by the
only mini-halos significantly abundant at those $z$. Only after $z\sim
50$ is the gas cold enough for it to fall into the potential wells of
reasonably abundant halos with $M\H\sim 10^5$ \modotd, giving rise to
the formation of the first generation stars.

\subsection{Unbound IGM}\label{nt}

Luminous sources reionize and reheat the diffuse unbound IGM,
hereafter simply the IGM. UV photons, with short mean free paths,
ionized bubbles around them which grow and progressively
overlap. Inside these bubbles, subbubbles with doubly ionized helium
develop due to the smaller fraction of more energetic UV
photons. X-ray photons with a much larger mean free path give rise
instead to a uniform background also heating the IGM by Compton
scattering (and through secondary ionizations, neglected in AMIGA).

Some amount of the diffuse gas in the IGM is accreted by massive
enough halos (inflows) or expelled from them (outflows). The gas mass
inflow rate, $\dot M\hg\in$, is proportional to $\dot M\H$
(Sec.~\ref{DM}), with proportionality factor equal to the current
baryon mass fraction in the IGM, calculated from the original total
baryon fraction taking into account the gas gains and losses into and
from halos. Gas outflows are triggered by supernova- (SN-) or active
galactic nucleus- (AGN-) driven winds (Secs.~\ref{reheating} and
\ref{AGNf}, respectively) when they cause the hot gas in the halo to
become unbound. Its typical rate, $\dot M\hg\out$, is taken equal to
the hot gas mass lost over the wind duration. Outflows from halos also
cause the chemical enrichment of the unbound IGM. As this effect takes
place only at the vicinity of halos hosting luminous sources, AMIGA
assumes that the metal pollution of IGM affects ionized bubbles only.

AMIGA deals with the properties of neutral and ionized IGM,
separately, distinguishing between \HeII and \HeIII ionized
regions. The evolution with $z$ of the IGM temperature, $T\IGM$, or,
more exactly, the average temperature $T\jj$ of the gas in phases j
$=$ I, II and III, for the neutral, singly, and doubly ionized
regions, respectively, is according to the differential equation (MSS)
\beq
\frac{\der\ln T\jj}{\der\ln(1+z)}=
2+\frac{\der \ln (\mu\jj \varepsilon\jj/n\jj)}{\der\ln(1+z)}\,,
\label{Tigm}
\eeq
where $\mu\jj$, $\varepsilon\jj$, and $n\jj$ are the average mean
molecular weight, energy density, and baryon density, respectively, in
region j. In equation (\ref{Tigm}), the term equal to 2 on the right
gives the cosmological adiabatic cooling, and the second term includes
Compton heating--cooling from the CMB and X-rays, heating--cooling by
ionization--recombination of the various chemical species, collisional
cooling of hot neutral regions, achievement of energy equipartition of
newly ionized--recombined material, inflows--outflows from halos, and
cooling by collisional ionization and excitation. The heating--cooling
by gravitational compression--expansion of density fluctuations
vanishes, neglecting non-linear effects, after averaging over each
ionization phase (MSS).

\begin{figure}
\vspace{-0.2cm}
\centerline{\psfig{file=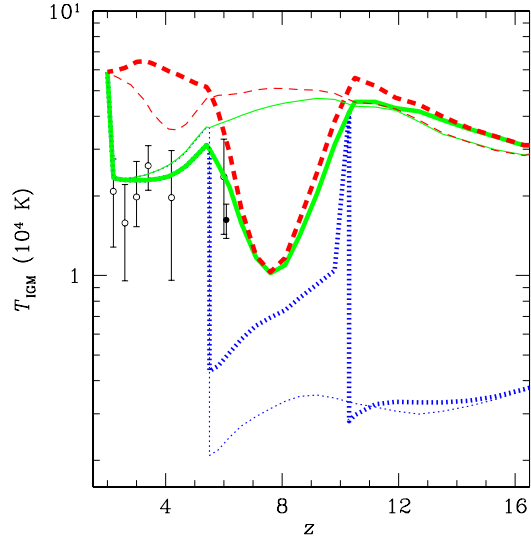,width=.48\textwidth,angle=0}}
\vspace{-.5cm}
\caption{Average IGM temperatures in neutral (blue dotted lines),
  singly ionized (green solid lines), and doubly ionized (red dashed
  lines) regions obtained from two models with identical values of all
  the parameters but those ($f_2$ and $f_3$; see below) characterizing
  the \pIII star IMF. The top-heaviest IMF (highest $f_2+f_3$ value)
  leads to double \HI reionization, at $z=10.3$ and 5.5 (thick lines),
  while the less top-heavy one (lowest $f_2+f_3$ value) leads to
  single \HI reionization at $z=5.5$ (thin lines). In both cases,
  there is one single \HeII reionization at $z=2$. Note the vertical
  jumps from a lower temperature to a higher one at the redshift of
  reionization of a given state (with the lower temperature) due to
  the fact it completely disappears. Indeed, in case of an eventual
  recombination period, the temperature of that state starts to evolve
  again from the value corresponding to the higher ionization
  state. Caution must be paid to the fact that, for the temperature of
  neutral regions not to be out of range, it has been shifted upwards
  by a factor 500 (in the case of double \HI reionization, only at $z
  > 10.3$). The observational measures of the IGM temperature (circles
  with error bars) refer to singly ionized regions.}
\label{temp}
\end{figure}

Figure \ref{temp} illustrates the kind of IGM temperature evolution
that can be obtained depending on the values of the free parameters of
AMIGA, such as $b_{\rm cl}$ (see below for the rest). The specific
solutions shown correspond to two distinct \pIII star initial mass
functions (IMF): the top-heavier one leads to double \HI reionization,
while the less top-heavy one leads to single reionization. As can be
seen, in the case of double reionization, there is a marked dip in the
temperature of the (singly and doubly) ionized IGM over the redshift
interval between the two full ionizations (at $z=5.5$ and 10.3),
absent in the case of single reionization. This is due to the drop in
the flux of ionizing photons at first complete ionization ($z=10.3$),
when \pIII stars stop forming because of the lack of neutral
regions. This causes a short period of \HI recombination until the UV
flux from normal galaxies becomes (at $z=8$) high enough for
reionization to start gain.

\subsection{Trapped Hot Gas}\label{cooling}

\subsubsection{Structure}

The hot gas in equilibrium within halos is not assumed to be
isothermal as often done in SAMs, but with a polytropic equation of
state with index $\Gamma=1.2$, consistently with the MEPS formalism
for halo growth. The gas that is accreted by a halo is shock-heated
and deposited at the instantaneous virial radius of the halo, meaning
that its spatial distribution grows from the inside out as that of the
DM \citep{Sea12a} with the only difference that the gas has a
polytropic equation of state as a result of the shock, whereas dark
matter follows a density profile \`a la NFW \citep{NFW97} set by the
rate at which halo accretes the non-collisional DM. It is the
preservation of the ratio of {\it total} energies between the gaseous
and dark components that fixes the value $\sim 1.2$ of $\Gamma$
\citep{Soea05}. Although this reasoning applies to halos grown by pure
accretion, the same result holds for halos having suffered major
mergers \citep{Sea12a,SS12b}. Such a gaseous structure is not only
expected on theoretical grounds, but it is also supported by
observations \citep{PSF03,Pea10} and simulations
\citep{VKB05,Shoea10}. Furthermore, it leads to X-ray scaling
relations from galaxy groups to rich clusters that are in very good
agreement with observation \citep{Soea05,BOV09}. Thus, the inner
structure of the hot gas within halos is also calculated directly from
the cosmology considered with no free parameters.

\subsubsection{Cooling}\label{coolrate}

The hot gas radiates, cools, and contracts in a runaway process that
leads to the infall of cold gas to the the halo center. The treatment
of cooling carried out in AMIGA is the same as in conventional
SAMs. The cooling radius, $r\cool(t)$, encompassing the gas having had
time to cool since the formation of the halo is found by equating the
age of the halo, $t_a$, to the characteristic cooling time, given by
the ratio between the energy density and the emissivity, ${\cal
  E}\hg(r)/\dot {\cal E}\hg(r)$, of the hot gas. The cooling rate is
then
\begin{equation}
\dot M\cool = 4\pi\,r\cool^2\,\mu m\p n\hg[r\cool(t)]\,{\der
  r\cool\over \der t}\,,
\label{Mcool}
\end{equation}
where $m\p$ is the proton mass, $\mu$ is the mean molecular weight of
the hot gas, and $n\hg$ is its particle number density.

Star formation begins to proceed at a significant rate in mini-halos
with $T\g\sim 10^3$ K. At these temperatures, the gas is essentially
neutral and atomic cooling is not effective. The only way such a
primordial gas can radiate is by spontaneous emission of
roto-vibrational molecular levels excited by collisions of atoms with
\HH molecules (with a fraction as small as $x\sim 10^{-6}$). In more
massive halos with virial temperatures above $10^4$ K, \HH molecules
are dissociated by collisions with atoms, and the spontaneous emission
of atomic electronic levels excited by collisions of atoms with free
electrons is the dominant cooling mechanism. Then, the higher the
metal abundance, the more effective cooling is. The full casuistry
found is the following:

\noindent (i) If the metallicity is higher than a critical value
$Z\crit$, and the gas is ionized ($T\g>10^4$ K), atomic line cooling
takes place. The emission through metallic lines keeps on operating
when the gas cools below $10^4$ K or when the density increases to the
point that the gas becomes shielded to ionizing photons and
recombines. Molecular and dust emission then become active (see next
item), but the total cooling rate is limited by the initial atomic
rate.

\noindent (ii) If the metallicity is higher than $Z\crit$ and the gas
is not ionized (no stars in the halo), molecular cooling can proceed
by means of \HHp and many other molecules synthesized on dust grains
as well as by the emission of dust itself. The rate of this complex
cooling process is unknown and, contrarily to case (i), it is not
limited by the rate of any previous cooling mechanism. But this case
can be neglected because the high-metallicity is indicative that star
formation has taken place in the halo progenitors. 

\noindent (iii) If the metallicity is below $Z\crit$ and $T\g >10^4$
K, a sequence of two different processes takes place. As at such
temperatures \HH molecules are dissociated due to collisions with H
atoms, the first mechanism to operate is atomic cooling. However, once
$T\g$ drops below $10^4$ K, most \HII recombines and, as there are no
free electrons, atomic cooling halts. Then, the gas switches to
\HHp-cooling for a concentration of this molecule corresponding to
equilibrium \citep{OH02}.

\noindent (iv) If the metallicity is lower than $Z\crit$ but $T\g <
10^4$ K, atomic cooling is not efficient because there are essentially
no free electrons that can excite atoms (not only because the gas is
mostly recombined, but also because the remaining electrons do not
have enough energy to excite hydrogen atoms at those temperatures, the
lowest excitation level requiring an energy of $1.2\times 10^5$ K), and
the gas cools directly by \HHp-molecular emission.

The latter two cases presume of course that the gas is not ionized by
luminous sources. Otherwise molecules could not form and molecular
cooling would not be effective. The critical metallicity, $Z\crit$, is
taken equal to $10^{-4}~Z_\odot$ \citep{SS05,STSON09,SO10}.

In case (i), the emissivity leading to equation (\ref{Mcool}) is given
by the usual expression for atomic cooling
\begin{equation}
\dot {\cal E}\hg(r)=n\g^2(r)\Lambda[T\g(r),Z\g]\,,
\label{dotE}
\end{equation}
with the cooling function $\Lambda[T\g(r),Z\g]$ drawn from
\citet{SD93}. 

In cases (iii) and (iv), the cooling rate depends on the abundance of
\HH molecules, and the situation is more complex owing to the strong
feedback that stars have on the \HH concentration. In metal-free
gases, this molecule forms through reactions catalyzed either by
electrons or by protons. See Table \ref{reacrates} for the different
possible reactions: the second and third channels for electrons, and
the 4th and 5th ones for protons. The former of these channels is the
most efficient although for completeness both channels are included in
AMIGA. The first reaction corresponds to recombination. We will come
back to recombination later on.

The concentration of \HH in the gas of a newly born halo is computed
in AMIGA according to the reactions and corresponding rates given in
the second column of Table \ref{reacrates} (taken from
\citealt{HSTC02}) for the appropriate density and temperature of the
gas in the halo and from the initial concentrations and total
abundances of all initial chemical species, namely, H, H$^+$, H$^-$,
\HH, H$_2^+$ and e$^-$, previously calculated for each progenitor,
starting from the trivial initial concentrations outside the
interpolation grid of AMIGA given by \citet{GP98}. 

Provided there is no star in the halo (otherwise molecular cooling is
inhibited), the gas cools efficiently and contracts until $n\g$ and
$T\g$ reach some critical values $n\c$ and $T\c$, respectively equal
to $10^4$ cm$^{-3}$ and $100~\rm K$. Then cooling halts. The
\HHp-emissivity leading to such a stable state is 
\begin{equation}
\dot{\cal E}\hg(r)=f\nHH(r)\,n\g^2(r)\,\Lambda\nHH[T\g(r)]\,,
\label{emis:cool}
\end{equation}
where $f\nHH$ is the number fraction of \HH molecules, and
$\Lambda\nHH(T\g)$ is the associated cooling function, given by
\citet{GP98}, for the \HH concentration at equilibrium (case iii) or
calculated in the way explained above (case iv). After reaching the
minimum temperature $T\c$, the cold gas accumulates at the halo center
until one Bonner-Ebert, or simply one Jeans mass
\begin{equation}
M\J=\left( \frac{\gamma \pi \kb T\c}{G \mu m\p} \right)^{3/2}\rho\c^{-1/2}\,,
\label{Jeans}
\end{equation}
is reached. In equation (\ref{Jeans}), $\gamma$ is the adiabatic
index, $\kb$ is the Boltzmann constant, $\rho\c$ is the mean inner
density of the isothermal sphere with temperature $T\c$ or, more
exactly, with temperature equal to the maximum between $T\c$ and the
CMB temperature at that $z$, as the cold gas is heated by the
background radiation. Then it fragments and collapses to form a small
cluster of metal-free stars of about 1000 \modotd.\footnote{The exact
  mass of these star clusters depends on $z$ owing to the fact that
  the temperature $T\c$ of the cold gas in the disk is bounded by the
  CMB temperature.}

\begin{table}
\caption{Rates for the reactions involved in \HH formation. $T$ 
is the gas temperature in K and $T_n = T/10^n$.}
\begin{tabular}{ll}
\hline
\hline
Reaction &Rate Coefficient (cm$^3$ s$^{-1}$)\\
\hline
H$^+$ + e$^- \rightarrow$ H + $h\nu$ & $k_1 = 8.4\times 10^{-11}T_3^{0.2}/
\left(T^{0.5}\left[ 1+T_6^{0.7} \right]\right)$\\
H + e$^- \rightarrow$ H$^- + h\nu$ & $k_2 = 1.4\times 10^{-18}T^{0.928}
e^{-T/[1.62\times 10^4]}$\\
H + H$^- \rightarrow$ \HH + e$^-$ & $k_3 = 1.3\times 10^{-9}$\\
H + H$^+ \rightarrow$ H$_2^+ + h\nu$ & $k_4 = 2.1\times 10^{-23}T^{1.8}
e^{-20/T}$\\
H + H$_2^+ \rightarrow$ \HH + H$^+$ & $k_5 = 6.4\times 10^{-10}$\\
\hline
\end{tabular}
\label{reacrates}
\end{table}         

All physical processes calculated so far, directly from the cosmology
considered with no free parameters, are consistent with both
observation and simulations.

\section{Stars}\label{stars}

\subsection{Star Formation}\label{starform}

Because of the presence of metals, case (i) leads to the formation of
ordinary Population I and II (\pIaIIf) stars, whereas cases (iii) and
(iv) to \pIII stars.  

In the metal-rich case, the cold gas collected at the halo center
tends to settle down in a centrifugally supported disk or it directly
feeds a central spheroid (see Sec.~\ref{galaxies}), where it gives rise
to star formation.

The star formation rate (SFR), $\dot M_{\rm sf, C}$, in the galactic
spheroid (C=B) or disk (C=D) is taken according to the usual
Schmidt-Kennicut law \citep{Kenn98},
\begin{equation}
\dot M_{\rm sf, C}=\alpha\G \frac{M_{\rm cg,C}}{{\tau\dync}}.
\label{sfr}
\end{equation}
where $M_{\rm cg,C}$ is the mass of cold gas available, ${\tau\dync}$
is the dynamical timescale at the half-mass radius of the component C
(see Sec.~\ref{mom}), and $\alpha\G$ is the star formation efficiency,
taken as a free parameter.

\pIaII stars form according to the IMF, $\phi(m)$, along the zero-age
main sequence and evolve along the respective mass- (and metallicity-)
dependent evolutionary tracks.\footnote{Caution must be paid to the
  definition of $\phi(m)$. Its integral is not normalized to unity
  (i.e. $\phi(m) \der m$ is not the fraction of stars with masses
  between $m$ and $m+\der m$); it is the integral of $m \phi(m)$ which
  is so normalized (i.e. $m\phi(m)\der m$ is the mass fraction in
  stars with such masses).} AMIGA is ready for any wanted IMF, but the
default one is the modified Salpeter IMF proposed by \citet{WTH08},
consistent with observations of the local IMF, which is characterized
by a power-law index equal to $-1$ in the range $0.1 < m/$\modot
$<0.5$, and equal to the Salpeter value ($-2.35$) in the range $0.5 <
m $/\modot $ <100$. This does not preclude, of course, that the IMF of
metal-poor stars is much more top-heavy (i.e. with a greater lower
mass or a less steep logarithmic slope).

\pIII stars are believed, indeed, to reach masses well above 100
\modotd. Those with $m \la 130$ \modot would explode as type II SNe
and produce metals typically according to the yield $p$ of ordinary
\pIaII stars (see Sec.~\ref{metals}), whereas those with $130 \la
m/$\modotd$\la 260$ would explode as Pair Instability SNe (PISNe) and
release about half their mass as metals. Lastly, those with $m\ga 260$
\modot would collapse into a black hole (BH) and leave no yield at all
\citep{HW02}. Thus, denoting the mass fractions of metal-free stars in
these three mass ranges of increasing mass as $f_1$, $f_2$, and $f_3$,
we have that parameters $f_2$ and $f_3$ ($f_1=1-f_2-f_3$) are enough
to characterize the \pIII star IMF and evolution.

\subsection{Stellar SEDs}\label{SEDs}

To calculate the emission from normal galaxies or low-mass ($m < 130$ \modotd)
\pIII stars an SED is assigned to each group of stars in the
color-magnitude diagram according to its time-varying spectral type
and luminosity class (see Sec.~\ref{galprop} for details). The
spectrum of massive ($m > 130$ \modotd) \pIII stars is approximated by that of
a black-body with effective temperature $T\eff$ equal to $\sim 10^5$ K
regardless of their mass \citep{BKL01}. The superposition of all these
spectra gives rise to the synthetic SED of the whole stellar
population of the galactic component under consideration as a function
of time.

The contribution $F_\lambda (t)$ to the galactic flux at wavelength
$\lambda$ at time $t$ of a given stellar population is
\begin{equation}
F_\lambda (t)= \int^t_0 \der t' \int_{m=0}^{\infty} \der
m \, \dot
M\sff(t-t') \,\phi(m)\, f_\lambda(m,t',Z\s)\,,
\label{sam:phot:eq}
\end{equation}
where $f_\lambda(m,t',Z\s)$ is the flux at wavelength $\lambda$ of one
individual star with initial mass $m$, initial metallicity $Z\s$, and
age $t'=t-t\s$, being $t\s$ its formation time, with
origin at the zero-age main sequence (i.e. $f_\lambda(m,t',Z\s) = 0$ for
$t'$ greater than the lifetime of the star), provided by the
adopted stellar population synthesis model (\sec \ref{galprop}).

\subsection{Stellar Feedback}

Stars affect the surrounding IGM in three different ways: by
increasing its metallicity through SNe and stellar mass losses, by
reheating it mechanically through SN shocks, and through radiative
losses. Since the richer in metals, the more easily the hot gas cools,
metal enrichment is a positive feedback for star formation. On the
contrary, reheating by SNe may cause part of the metal-enriched
interstellar medium (ISM) to escape from the halo (outflows) and,
consequently, it is a negative feedback like photo-dissociation of
molecules, and photo-ionization and reheating of the IGM. In all these
feedback processes there are important differences between \pIaII and
\pIII stars.

\subsubsection{Photo-dissociation}\label{diss} 

\HH is dissociated by photons with energy below the Lyman
limit, in the so-called Lyman-Werner (11.28 -- 13.6 eV) bands \citep{HAR00}. 
This effect operates when SFR becomes intense enough for
a cosmic soft UV background to build up.

The rate at which dissociation takes place can be approximated by
\citep{AAZN97}
\begin{equation}
k\diss=1.38\times10^{-12}J_{21}(h\nu=12.87\, eV),
\end{equation}
in s$^{-1}$, being $J_{21}$ the flux in units of $10^{-21}~\rm{erg}$
$\rm{s}^{-1}~\rm{cm}^{-2}$ $\rm{Hz}^{-1}~\rm{str}^{-1}$.  This flux
should be essentially homogeneous and isotropic since the distance
traveled by the photons is far larger than the mean separation
between halos (absorption by the medium can be neglected). In these
conditions, redshifting of the photons must be taken into
account. Thus, by integrating the flux of dissociating photons from
all \pIII stars in a given volume one can obtain the emissivity of
dissociating photons at each $z$, $j_\nu(z)$, and from it the
corresponding flux
\begin{equation}
J_\nu(z)=\int_z^{z\maxi}dz'c\,\frac{\der t}{dz'}j_{\nu'}(z'),
\label{jnu}
\end{equation}
where $c$ is the speed of light, and $z\maxi=(13.6/11.28)(1+z)-1$ is
the redshift at which photons with an energy of 13.6 eV are redshifted
to 11.28 eV. Note the upper bound in the integral of equation
(\ref{jnu}) coming from the fact that any UV photon emitted at a
redshift $z\emi>z\maxi$ falling into the soft UV bands after
redshifting to $z$ will have been previously absorbed by the neutral
IGM. Actually, the flux given by equation (\ref{jnu}) is shielded
inside halos due to the molecules produced since virialization.  Thus,
to calculate the photo-dissociation rate the emissivity given in
equation (\ref{jnu}) must be multiplied by the shielding factor
$F\shi$, estimated through \citep{DB96}
\begin{equation}
F\shi={\rm min}\left[1,\left(\frac{N\nHH}{10^{14}\,\rm{cm}^
{-2}}\right)^{-3/4}\right],
\end{equation}
where $N\nHH$ is the column \HH density. 

\subsubsection{Photo-ionization}\label{reion} 

To calculate the flux of \HIp/\HeIp- and \HeIIp-ionizing photons
emitted by \pIII and \pIaII stars (as well as AGN) one must consider
the different SEDs of the emitting objects. That of metal-rich and
low-mass metal-poor stars is provided by the stellar population
synthesis model, taking into account the star formation and
metallicity histories of the emitting populations. The flux of
ionizing photons from zero-metallicity \pIII stars and the nebular
emission they induce is computed according to \citet{Schaerer02}, for
the particular IMF (i.e. the mass fractions $f_1$, $f_2$, and $f_3$)
considered.

At each $z$, we compute the flux of ionizing photons escaping from
galaxies in halos with different masses along the whole (relevant)
range, and integrate for the halo MF corresponding to that $z$ in the
ionized environment. For halos with virial temperatures lower than
$10^4$ K, the escape fraction of photons above the Lyman continuum
limit, $Ly\c$, is obtained by subtracting those photons captured by
the neutral gas present in it. While if the temperature is higher than
$10^4$ K, we assume some escape fraction, $f\esc$. The possibility
that $f\esc$ increases with increasing $z$ has been considered in
order to obtain a reionization at high-$z$ as suggested by the
analysis of CMB anisotropies \citep{P09,KFG12,A12}. However, such a
behavior is not supported by observation, so we adopt in AMIGA the
usual more conservative assumption of a constant $f\esc$, taken as a
free parameter.  Recombination, both inside and outside halos where
ionizing photons are produced, is also taken into account as it leads
to the absorption of more ionizing photons.

The evolving \HII and \HeIII volume filling factors, $Q\nHII$ and
$Q\nHeIII$, are governed by the differential equations for trivial
initial conditions at the dark ages (MSS)
\beq 
\frac{\der Q\Si}{\der t}\!=\!\frac{\lav{\dot
    N}\Si\rav}{\lav n_{\rm
    S}\rav}
\!-\!\left[\!\left\lav\!
  \frac{\alpha_{\rm SI}(T\IGM)}{\mu\se}\!\right\rav\bSi \!\!\frac{C\,\lav
    n\b\rav }{a^3(t)}\!+\!\frac{\der \ln \lav n_{\rm S}\rav}{\der
    t}\right]\!Q\Si\,,
\label{rec}
\eeq 
where subscripts S, SI, and SII stand for H, \HIp, and \HIIp, or
\HeIp, \HeIIp, and \HeIIIp, and angular brackets mean averages over
the regions denoted by subscript (in the lack of any subscript the
average is over the whole IGM). The average of a function $f(T\IGM)$
of the IGM temperature in the region j is taken equal to
$f(T\jj)+(\der^2 f/\der T^2){{\sigma\T^2}\jj}/2$, with $T\jj$ equal to
the mean temperature in that region and ${\sigma\T^2}\jj$ the
corresponding variance. In equation (\ref{rec}), $\lav n\b\rav$ is the
comoving cosmic baryon density, $a(t)$ is the cosmic scale factor,
$\mu\se$ is the electronic contribution to the mean molecular weight,
$\dot N\Si$ is the comoving metagalactic ionizing photon rate density
due to luminous sources and recombinations (calculated according to
\citealt{Me09}) to \HeII and \HeI ground states for \HIp-ionizing
photons (for simplicity, the contribution from \HeII
\lya\ recombinations is neglected), $\alpha_{\rm SI}$ is the
recombination coefficient to the SI species, and $C$ is the ionized
clumping factor.

Cosmological $N$-body simulations allow one to estimate the clumping
factor from the observed fluctuations in DM, $C_\rho$, for which we
have practical analytic fits \citep{IEtal07,RT11}. Did baryons trace
mass, $C_\rho$ and $C$ would be identical. Unfortunately, the limits
of the diffuse IGM are hard to established in terms of the DM density
field. On the other hand, the increased pressure in the ionized gas
may largely reduce its density fluctuations
\citep{MEtal00,P09,F12}. For this reason we adopt the relation
$C=b_{\rm cl} C_\rho$, with the matter clumping factor $C_\rho$ drawn
from simulations and the bias factor $b_{\rm cl}$ taken as a free
parameter.

\subsubsection{SN Reheating}\label{reheating} 

The secular effects of reheating on the unbound IGM are accounted for
through equations (\ref{Tigm}), while the extra energy of
non-gravitational origin imparted to the hot gas trapped in halos is
transferred jointly with the gas itself (and galaxies) to the
descendants of every halo.

X-ray photons produced in SNe (from free-free emission and inverse
Compton scattering of CMB photons by relativistic electrons) and, at a
lesser extent, emitted from very massive \pIII stars, ordinary binary
stars, and AGN, Compton heat the IGM and increase the entropy-floor of
the non-trapped gas. The fraction of the SN energy converted to X-rays
is about 1\% \citep{OH03}.

SNe also reheat mechanically the ISM in disks and spheroids of normal
galaxies as well as the hot gas in halos with metal-poor galaxies. For
a given stellar mass at formation, $M\sff$, some fraction is in
massive fast-evolving stars that quickly explode as SNe. When a SN
occurs, some amount of the energy released is imparted to the
surrounding ISM, causing part of it to blow off to the halo, in case
of ordinary stars, or it is directly imparted to the hot gas in the
halo, which can be ejected from it (usually it is), in the case of
very massive metal-free stars.

The condition that the reheated ISM leaves the component C of galaxies
and joints the hot gas in the halo leads to the usual expression,
\begin{equation}
M_{\rm rh,C}=\epsilon\C \frac{2\,\eta\SN E\SN}{V\C^2-V\hg^2} M_{\rm sf,C}\,
\label{m:reh}
\end{equation}
for the mass of reheated gas, its time-derivative leading to the rate,
$\dot M\rhc$, at which the ISM of component C is reheated and expelled
from the galaxy. In equation (\ref{m:reh}), $E\SN\approx 10^{51}$ erg
is the energy produced in one Type II SN, $\eta\SN=0.0144$ is the
number of SN explosions per solar mass unit over the typical duration
of a starburst (0.2 Gyr) of stars formed instantaneously with the
(modified) Salpeter IMF, $V\hg$ is the thermal velocity of the hot gas
at the halo half-mass radius, $V\C$ is the circular velocity at the
radius used to set the typical dynamical time of component C, and
$\epsilon\C$ is the corresponding SN reheating efficiency.

Hydrodynamic studies indicate that reheating of ISM by type II SNe
triggers galactic winds only in spheroids, which is also in agreement
with observation. The reason for this would be that, in disks, only a
small fraction of the SN energy is directed towards the plane where
the ISM lies, which greatly diminishes the heating efficiency. For
this reason, we take $\epsilon\C$ equal to one for C=B
(e.g. \citealt{DS86,MF99,SS00}), and equal to zero for C=D. However,
when the gas has little angular momentum and the stable disk is found
to be smaller than the corresponding spheroid (bulge), we assume it
settles in an oblate pseudo-bulge rather than in a thin disk, so
$\epsilon\C$ is then also taken equal to one. Of course, the effective
amount of reheated gas leaving a galactic component depends not only
on the reheating efficiency but also on the gravitational pull of the
galaxy, accounted for through the circular velocity $V\C$.

In the case of \pIII star clusters, equation (\ref{m:reh}) also holds
but with $V\hg=0$ and the circular velocity $V\C$ at the half-mass
radius of the halo as the reheated gas is then expelled from it out to
the unbound IGM. On the other hand, the energy $E\SN$ liberated by one
PISN explosion is two orders of magnitude larger than in Type II SNe
\citep{FWH01,HW02}, and the expected number $\eta\SN$ of SN explosions
per solar mass unit of stars formed in a typical starburst is $0.0015$
\citep{Schaerer02}. The reheating efficiency of PISNe explosions is
likely also different from that of normal SNe, but the exact value
does not matter provided it is large enough for mini-halos to lose the
hot gas in those explosions. This is indeed what happens for
$\epsilon$ equal to unity as also adopted for simplicity in AMIGA.

The stellar mass loss, $M\lossc$, going into the ISM of component C
from stars with masses spanning from $m_1$ to $m_2$ in a stellar
population with total mass $M\sfc$ at formation is
\begin{equation}
M\lossc = M\sfc \int_{m_1}^{m_2} [m - w(m)]\; \phi(m)\;\der m\,,
\label{prev}
\end{equation}
where $w(m)$ is the mass of the remnant left after the star with $m$
dies. This expression can be readily extended in order to account for
the entire star formation history of a given stellar population. This
leads to the following stellar mass loss rate,
\begin{equation}
\label{eqn:yield}
\dot M\lossc = \int_{0}^{m\up} \dot
M\sfc[t-\tau(m)]\; [m - w(m)]\; \phi(m)\; \der
m\,,
\label{preira}
\end{equation}
where $\tau(m)$ is the lifetime of stars with $m$, and $m\up$ is the
IMF upper mass.

\subsubsection{Metal Enrichment}\label{metals} 

The amount of metals ejected by stars into the ISM over their life
and when they die as SN explosions depends on whether they are
metal-rich or metal-poor. As mentioned, the yield of \pIII stars
depends on their initial mass. According to the definition of $f_1$,
$f_2$, and $f_3$ (\sec \ref{starform}), the mass fraction in massive
($m> 130$ \modotd) \pIII stars ending up locked into BHs is
\beq 
\beta\III=f_3\,,
\label{mu}
\eeq 
while their yield is
\begin{equation}
p_{\rm III}=0.5\,f_2\,.
\label{y3}
\end{equation}

Once the hot gas metallicity reaches the value $Z\crit$, the \HH
density is no longer relevant as atomic cooling becomes the most
efficient cooling mechanism. Then, Pop II stars begin to form in disks
and spheroids.  \pIaII stars and the less massive \pIII stars liberate
metals by type II SN explosions and, at a lesser extent, throughout
their life. In AMIGA, we follow the mass loss of \pIaII stars
according to their specific evolution and compute the mass of metals
they eject, supposed to mix up with the cold ISM under the
instantaneous recycling approximation, IRA (\citealt{T80}).

Equation (\ref{eqn:yield}) can also be readily adapted for the
computation of the metal mass gain by ISM owing to stellar
evolution by means of the substitution
\begin{equation}
m - w(m) \longrightarrow p(m)\,m\,,
\end{equation}
where $p(m)$ is the yield of stars with $m$, which we can approximate
by the global average value $p$, taken equal to 0.03,\footnote{The
  theoretical value of $p$ for the (modified) Salpeter IMF is $\sim
  0.02/(1-R)$ with the recycling fraction $R$ equal to about $0.4$
  (\citealt{CLBF00}; see also \citealt{Mea07}).}. Thus, neglecting,
according to IRA, the lifetime of massive stars, those essentially
contributing to the chemical enrichment of the ISM, we arrive at the
following metal mass loss rate into the ISM of component C due to
stellar evolution
\begin{equation}
\dot{(ZM)}\lossc=  p\,\dot M\sfc \!\!\int^{m\up}_{m\IRA}
\!\!\der m\,m\,\phi(m)\equiv p\,\PhiIRA\dot M\sfc\,,
\label{postira}
\end{equation}
where $m\IRA= 10$ \modot is the effective lower mass of stars producing
metals, and $\PhiIRA$ is equal to $0.1$ for the (modified) Salpeter
IMF.

\section{Galaxies}\label{galaxies}

\subsection{Inner Structure}\label{mom}

The disk stability condition used in AMIGA is the simple global one
provided by \citet{VdB98}. The shape of the disk of the central galaxy
is computed self-consistently through the iterative procedure
described in \citet{MMW98} from the specific angular momentum of the
gas at the cooling radius in the halo, taken equal to that of dark
matter distributed according to the results of $N$-body simulations
(e.g. \citealt{CT96,Bull01}). This completely determines the scale
length $r\d$ or, equivalently, the central surface density,
$\Sigma(0)=M\d/(2\pi\ 0.83\ r\d^2)$, of the exponential disk with
total mass equal to $M\d$. Hence, the disk structure is also set
without introducing any free parameter. When a central disk galaxy
becomes a satellite (see Sec.~6.2) it conserves its shape.

If the disk is unstable or its stability radius is smaller than the
spheroid radius, the cold gas coming from the halo directly goes into
the spheroid. As the gas reaching the spheroid is the first to cool,
it contributes with a very low angular momentum to the spheroid, which
is for simplicity neglected. Some cold gas also reaches the central
galaxy through captured satellites (see Sec.~\ref{inter}). The orbital
momentum of satellites is assumed to be random so that such captures
do not to alter (in average) the angular momentum of the disk.

Owing to the lack of analytic treatment for violent relaxation,
spheroids are the only systems whose inner structure cannot be
causally linked to cosmic properties set by cosmology. AMIGA assumes
them with 3D density profiles of the \citet{Hern90} form (whose
projection in 2D approximates the $r^{1/4}$ law) with scale length
$r\B=r\e/1.81$, where $r\e$ is the effective (half-mass) radius of the
2D profile. Spheroids forming with no gas and, hence, suffering no
dissipative contraction, are assumed to satisfy the relation
$r\e\approx A M\B^{\gamma\B}$ \`a la \citet{Korm77} between the
effective radius $r\e$ and total stellar mass $M\B$, with constants
$A$ and $\gamma\B$ such to recover the observed values of $r\e$ of
local spheroids with extreme masses \citep{ShETal03,Guz97}.\footnote{
  Current spheroids with the highest and lowest masses should have
  suffered no dissipative contraction, indeed, because the respective
  initial values of $\rho\cg$ and $Z\cg$ are very small (see
  eq.~[\ref{radius2}] below).}  While those forming with some amount
of gas suffer, during star formation, dissipative contraction from the
previous initial configuration. In the Appendix, we derive the
following physically motivated differential equation for the
dissipative contraction of the scale radius,
\beq 
r\B^2(t)\,\frac{\der r\B}{\der t}= 
-\frac{Z\cgb^{1/2}(t)M\cgb(t)} {Z_\odot^{1/2}\rho\dis\tilde\tau\acc},
\label{radius2}
\eeq 
where $\tilde\tau\acc$ is the universal time elapsed since the
formation of the spheroid to the quenching of star formation due to
the action of the AGN, and $\rho\dis$ is a critical dissipation
density, taken as a free parameter. When contraction is so marked that
the density of the cold gas reaches the typical density ($10^6$
particles cm$^{-3}$) of dense molecular cores in local galaxies, AMIGA
assumes that the gas cloud fragments to form stars without suffering
any further contraction.

\subsection{Galaxy Interactions}\label{inter}

As halos merge and accrete, they accumulate more and more galaxies. In
a halo merger or in the accretion of a halo by a more massive one, the
most massive galaxy becomes the new central galaxy, and all the
remaining galaxies become its satellites. When a central galaxy
becomes a satellite, its original halo is truncated and part of the
dark matter remains bound to it with the original mass
distribution. The truncation radius is taken equal to the minimum
between the original halo radius and two times the maximum optical
radius (i.e. the radius encompassing 0.83 the total mass) of the disk
and the spheroid, any choice between one and three times that value
leading to almost indistinguishable results.

After the formation of a new halo at a major merger, the radial
location of all satellite galaxies is reconstructed according to the
PDF arising from the halo density profile. When a halo is accreted,
all its galaxies are located at the instantaneous radius of the
accreting halo according to its inside-out growth during accretion
(Sec.~\ref{DM}). The velocities of satellites are also normally
distributed in bins of velocity modulus and pitch-angle according to
the respective PDFs, at the satellite radius, derived from the halo
velocity dispersion and anisotropy profiles according to
\citet{Sea12a} and \citet{SS12b}, in agreement with the results of
simulations.

Going through all bins of initial conditions, we determine the time of
orbital decay (by dynamical friction) of the satellites according to
the prescription by \citet{GMS94}. This informs us on the expectation
number of captures and capture times, and the ending radial
distribution of the surviving satellites. After sorting the capture
times of all satellites, we follow the growth of the central galaxy by
accretion of new cold gas between consecutive captures, then we
compute the change of the galaxy properties owing to the new satellite
capture. Following the usual procedure in SAMs, AMIGA assumes that,
when the ratio between the masses of the satellite and the central
galaxy is larger than $\Delta_{\rm m}$, the capture is a merger with
destruction of the central galaxy giving rise to a spheroid. Otherwise
the gas of the satellite is incorporated to the disk (if any) and
stars to the spheroid of the central galaxy without destroying it. If
the central galaxy is a spheroid both the gas and stars of the
captured galaxy are deposited in the spheroid, causing a starburst and
the feeding of the central MBH (see Sec.~\ref{MBH}). On the contrary,
when new cold gas is incorporated to a stable disk, it causes the disk
to smoothly develop with continuous star formation. This results in a
variety of galaxy morphologies spanning from pure spheroids
(ellipticals) to galaxies with disks and bulges (disk galaxies). We
adopt $\Delta_{\rm m}=0.3$ so as to obtain a distribution of disk to
bulge luminosity ratios in agreement with observation of the nearby
universe \citep{SSS}.

Spiral galaxies moving inside halos suffer the effects of ram-pressure
from the hot gas. If it is strong enough according to the condition
given by \citet{GG72}, spirals lose all their ISM and the stripped,
chemically enriched, gas returns to the halo where it thermalizes and
mixes with the hot gas present there. For simplicity, AMIGA assumes
that the full recycling of the stripped gas is achieved when the halo
suffers a new merger or is accreted by a more massive halo.

Satellites can also lose mass into the intrahalo medium via tidal
interactions with other galaxies as they orbit inside the halo. The
typical mass loss rate due to tidal encounters is taken from
\citet{AW85}.  The mass lost includes dark matter as well as stars and
gas in the proportions found in the galaxy disk or spheroid. The only
satellites assumed to produce appreciable tides to a galaxy with a
given mass $M\G$ are those with masses equal to or greater than $\deli
M\G$, with $\deli$ a free parameter.

Interactions among satellites trigger non-axisymmetric perturbations
(bars and spiral arms) in the gaseous component of disks giving rise
to the transport of angular momentum outwards and the infall of
material to the bulge through bars.  The fraction of disk mass
transferred to the bulge is proportional to the strength of the
interaction, measured through the change in the orbital energy of the
galaxy in the impulsive approximation, with proportionality factor
$\chi\DB$ taken as a free parameter.

\section{Massive Black Holes}\label{MBH}

\subsection{MBH Feeding}\label{BHG}

MBHs are supposed to arise from the BH remnants of very massive ($m>
260$ \modotd) \pIII stars, which coalesce in one mini-MBH per star
cluster.  In galaxy mergers or captures, the MBHs of the progenitor
galaxies are assumed to migrate, by dynamical friction, to the center
of the new spheroid where they form a binary system. Binary MBHs break
if the recoil velocity produced by emission of gravitational waves (in
the Newtonian approximation; \citealt{Fit83}) is larger than the
galaxy escape velocity \citep{Bea11}, in which case the less massive
MBH escapes to the halo. Otherwise, the binary system quickly
coalesces \citep{SM07}.

MBHs also grow by accreting part of the gas that reaches their host
spheroid. Spheroids collect matter in three different ways: by means
of cooling flows of gas with low angular momentum, at wet mergers of
similarly massive galaxies, and via disk-to-bulge mass transfers. Part
of the gas loses angular momentum reaches the central region where
it feeds the central MBH.

Following \citet{Hea03}, AMIGA assumes that the gas mass accretion
curve onto the MBH, $M\BH^{\rm g}(t)$, scaled to the total accreted
mass, has a bell-shaped universal form with characteristic timescale
$\tau\acc$ (the AGN duty cycle; see below) equal to 0.1 Gyr. The only
exception is at the beginning of gas cooling after halo formation if
the central galaxy is a naked stellar spheroid. Then, the angular
momentum of the falling gas is very small and there is no hindrance
for the gas to directly reach the center of the main galaxy, so the
accretion rate into the central MBH is simply taken equal to the
cooling rate in the halo.

Apart from gas, MBHs accrete stars lying at the center of the spheroid
\citep{MM03} at the rate
\begin{equation}
\dot M\BH^{\rm s}=\alpha\BH\frac{M\sb}{\tau\BH}\,,
\end{equation}
where $\tau\BH$ is the dynamical time in the region of gravitational
influence of the MBH, with typical radius $r\BH$ equal to
$GM\BH/\sigma\B^2$, being $\sigma\B$ the stellar velocity dispersion
in the spheroid, and where $\alpha\BH$ is the MBH feeding
efficiency. In principle, $\alpha\BH$ should be taken as a free
parameter, but for all reasonable values tried, the resulting stellar
feeding is insignificant compared to the gas feeding, so we have taken
it simply equal to 0.01.

As a result of all these feeding mechanisms, MBHs grow at the center
of spheroids in such a way that they end up satisfying the observed
\citet{Mag98} relation between MBH masses and the stellar masses of
the host spheroids.

\subsection{AGN feedback}\label{AGNf} 

AMIGA assumes all AGN with the same typical intrinsic spectrum
independent of redshift. The continuum is described by two power laws,
crossing each other at a wavelength equal to 1100 \AA\ (big blue
bump). The optical spectral index in frequency (we define as
``optical'' the slope longwards of 1100 \AA) has a typical value of
0.5, and the UV index shortwards of 1100 \AA\ is equal to 1.76
\citep{WLZ98}. The most important emission lines (Ly$_{\alpha}$,
Ly$_{\beta}$, MgII, CIII, CIV, SiIV, H$_{\alpha}$, H$_{\beta}$ and
H$_{\gamma}$) and the small blue bump centered at $\sim$ 3000 \AA\,
are added to the above continuum with varying equivalent widths. From
such a spectrum and the bolometric luminosity of any given AGN,
inferred as explained in Section \ref{AGN}, one can readily compute
its rest-frame extinction-free flux of ionizing photons and the
associated energy.

AGN contribute to the X-ray background with 0.04 of their bolometric
luminosities \citep{VF07}. But the most important feedback of AGN is
the mechanical reheating of the gas inside galaxies. As mentioned,
when new gas reaches the spheroid a starburst takes place and the MBH
begins to accrete gas. At about half the increasing branch of the MBH
accretion curve, the gas reheated by the enlightened AGN begins to be
expelled back into the halo, which will ultimately cause the quenching
of the ongoing starburst \citep{Spring05} and the braking of gas
accretion onto the MBH. The total mass increase of the MBH is
estimated as the mass of gas remaining in the spheroid at the maximum
of the accretion curve minus the mass of gas reheated by the AGN and
expelled back into the halo, the reheating rate being given by
\begin{equation}
\dot M\rh^{\rm AGN}=\epsilon\AGN \frac{2\,L(t)\,c^{-2}}{V\B^2-V\hg^2}\,,
\label{m:rehAGN}
\end{equation}
where $\epsilon\AGN$ is the quasar-mode \citep{Bea06} AGN heating
efficiency, taken as a free parameter, and $L(t)$ is the AGN
bolometric luminosity (see Sec.~\ref{AGN}). Of course, AGN radiate at
most at the Eddington limit, so, in low mass MBHs, this reheating rate
may not be enough to expel all the gas remaining in the spheroid after
the feeding of the MBH. Then, the starburst continues, with the
dynamical timescale $\tau\dynb$ of the final contracted spheroid,
until all the gas is exhausted.

A second AGN feedback is the so-called radio-mode heating of the hot
intrahalo gas \citep{Crea06} taking place when the MBH lies within a
naked spheroid directly fed by cooling flows with small angular
momentum (see \ref{BHG}). In this case, about one tenth of the
bolometric AGN luminosity is transferred mechanically to the hot gas
in the halo through relativistic jets \citep{Crea06,Aea06}, which
slows down the cooling of the hot gas there, possibly even halting it
in the case of massive enough MBHs. Such a reheating is completely
determined by the amount of cold gas feeding the MBH and the AGN
radiation model described above, so it introduces no extra parameter.

Lastly, AGN also ionize and reheat the IGM outside halos. The escape
fraction of ionizing photons from AGN is taken equal to the
above mentioned escape fraction $f\esc$ of ionizing UV photons from
galaxies.

\section{Masses and Metallicities}\label{mass} 

As a consequence of all the preceding processes, baryons circulate
through the different phases and galactic components (see
Fig.~\ref{Defs}).

Specifically, in periods between sudden mass changes due to halo
mergers and galaxy captures and mergers, the masses of such phases and
components in any given halo evolve smoothly according to
the following set of differential equations
\begin{eqnarray}
{\der M\hg\over\der t}\!=\! \sum_{i=1}^N\!\Big\{\!\dot M\rh^{\rm
  AGN~(i)}\!+\!\!\!\sum_{\mathrm C=B,D} \!\!\left[\dot M\rhc\ui-\dot
  M\coolc\ui\right]\!\!\Big\}~~~~~~\nonumber\\ +\dot M\hg\in-\dot
M\hg\out~~~~~~~~~~~~~~~~~~~~~~~~~~~~~~~~~~~~~~~~\label{e1}\\ {\der
  M\cgc\ui\over \der t}\!=\!\dot M\coolc\ui\!-\!\dot M\rhc\ui\!-\!\dot
M\sfc\ui\!+\!\dot M\lossc\ui\!-\!\dot M\BHc\gi~~~~~\label{e2}\\ {\der
  M\sc\ui\over \der t}=\dot M\sfc\ui-\dot M\lossc\ui-\dot
M\BHc\si~~~~~~~~~~~~~~~~~ \label{e3} \\ {\der M\BH\ui\over\der t}=
\dot M\BH\gi+\dot M\BH\si\,,~~~~~~~~~~~~~~~~~~~~~
\label{e4}
\end{eqnarray}
where all the rates denoted by a dot on the right are known functions
of the evolving DM mass, $M\H$, provided by the EPS formalism, and the
hot gas, cold gas, stellar and MBH masses, $M\hg$, $M\cgc\ui$,
$M\sc\ui$, and $M\BH\ui$, whose evolution is being followed.

To render the notation in equations (\ref{e1})--(\ref{e4}) more
compact we have introduced the following definitions according to
whether galaxies are the central object or satellites, and the
galactic components are disks or spheroids: $\dot M\BHd\gi=\dot
M\BHd\si=\dot M\rhd^{\rm AGN~(i)}=0$, while $\dot M\BHb\gi=\dot
M\BH\gi$ and $\dot M\BHb\si=\dot M\BH\si$; in addition, $\dot
M\coolc\un1=0$, and either $\dot M\coolb\u1=0$, $\dot
M\coold\u1=\dot M\cool$, and $\dot M\sfb\u1=\dot M\rhb\u1=0$ or $\dot
M\coolb\u1=\dot M\cool$, $\dot M\coold\u1=0$, and $\dot M\sfd\u1=\dot
M\rhd\u1=0$, depending on whether or not star formation takes
place in a stable disk.

Those mass exchanges between phases are accompanied, of course, by
metal exchanges. As a consequence, in periods between sudden mass
changes due to captures and mergers, the metallicity of the hot gas,
cold gas, and stars, $Z\g$, $Z\cgc\ui$, and $Z\sc\ui$, in any given
halo with mass $M\H$, and of the ionized IGM associated with it,
$Z\IGM^{\rm H}$ (the metallicity $Z\IGM$ of the total ionized IGM is
the result of the metal losses by all halos lying in ionized regions)
evolve according to the set of equations
\begin{eqnarray}
{\der [Z\IGM^{\rm H} M\IGM^{\rm H}] \over \der t}=-\,Z\IGM \dot
M\hg\in+Z\hg \dot M\hg\out~~~~~~~~~~~\label{eZ}\\ {\der [Z\g M\g]
  \over \der t}\!=\!\sum_{\,i=1}^N \!\!\Big\{ Z\cgb\ui \dot M\rhb^{\rm
  AGN (i)}\!+\!\!\sum_{\mathrm C=\!B\!,D} \!\!\Big[Z\cgc\ui \dot
  M\rhc\ui~~~~~~\nonumber\\ - Z\g \dot M\coolc\ui\Big]\Big\} +Z\IGM
\dot M\hg\in-Z\hg\dot M\hg\out~~~~~~~~\label{eI}\\ {\der [Z\cgc\ui
    {M\cgc\ui}] \over \der t} = Z\g \dot M\coolc\ui-Z\cgc\ui \dot
M\rhc\ui~~~~~~~~~~~~~~~~\nonumber\\ +[p\,C\eff-
  Z\cgc\ui]\,\dot{M}\sfc\ui-Z\cgc\ui\dot
M\BHc\gi~~~~~~~~~~\label{eII}\\ {\der [Z\sc\ui {M\sfc\ui}]\over\der
  t} = Z\cgc\ui\,\dot M\sfc\ui-Z\sc\ui\dot M\BHc\si\,,
~~~~~~~~~~~\label{eIII}
\end{eqnarray}
coupled to the previous set (eqs.~[\ref{e1}]--[\ref{e4}]), where
$M\IGM^{\rm H}$ is the mass of that part of the ionized IGM associated
with the halo, equal to $M\H$ times the current baryon mass fraction in
ionized regions.

Strictly, equations (\ref{e1})--(\ref{eIII}) hold for halos harboring
normal galaxies. In the case of primordial \pIII star clusters, the
corresponding equations are somewhat different owing to the fact that
there is neither cooling in halos (stars photo-dissociate and even
photo-ionize the hot gas) nor cold gas in galaxies. Then, the mass and
metals lost by \pIII stars go directly into the hot gas, and the gas
reheated through PISN leaves the halo, liberating metals into the
surrounding IGM. In fact, these are essentially the only outflows from
halos opposed to the inflows mentioned in Section \ref{strat}, and
hence, the only vector for the metal enrichment of the IGM
\citep{RollEtal09,GGBK10,WTNA12}. Indeed, the gas ejected from normal
galaxies through type II SNe- and AGN-driven flows go into the halo
where it enriches the metallicity of the hot gas
(e.g. \citealt{SB10,Spring05}). In principle, it might also leave the
halo, but according to the value of $V\hg$ adopted in equations
(\ref{m:reh}) and (\ref{m:rehAGN}), the reheated gas leaves the
specific energy of the hot gas in the halo essentially
unaltered,\footnote{The only change is due to the cooling of its inner
  hottest fraction.}, so the possibility of those outflows is
actually ignored.

AMIGA also follows the detailed exchanges of carbon. The reason for
this is that carbon abundance is a more direct observable than 
metallicity $Z$, while the carbon mass fraction in the yields $p\III$
and $p$ of metal-poor and metal-rich stars are very different. For
such a monitoring, we adopt the carbon mass fraction in the two yields
provided by \citet{Schaerer02} and \citet{RPMZ09}.

When reheated gas (with increased metallicity) returns to the halo, it
takes some time to mix with the hot gas present there. In fact, during
the smooth evolution of a halo, viscosity causes the gas ejected from
the central spheroid (not from satellites) to stay stuck at the
cooling front, where it will be the next to cool.\footnote{Although
  the reheated gas does not reach the median halo radius, its higher
  specific energy will be transferred to the hot gas at the cooling
  radius which will expands and so on, until the whole hot gas is
  rearranged without any significant increase in its total specific
  energy.} The surviving reheated gas mixes with the hot gas when the
halo merges or is accreted. Specifically, in a merger, the gas
recently reheated is mixed with the hot gas lying in the inner $h\rec$
fraction. While, if the halo is accreted, the gas recently reheated is
mixed with the outer $1-h\rec$ fraction of the new halo. Only in next
major merger or accretion event is the surviving part of that reheated
gas definitely mixed with the hot gas in the new halo. The hot gas
recycling fraction $h\rec$ is a free parameter of AMIGA.

Therefore, although this is not reflected in equations
(\ref{e1})--(\ref{e4}), during periods of smooth evolution between
halo captures and mergers, the hot gas in halos is stored in two
separate compartments: the outer initial $1-h\rec$ fraction, where
the gas participates in inflows--outflows with the outer IGM, and the
inner initial $h\rec$ fraction, where it participates in
cooling--reheating exchanges with galaxies. When the inner
compartment is empty, cooling-reheating continues in the outer one.

\begin{figure}
\vspace{-0.7cm}
\centerline{\psfig{file=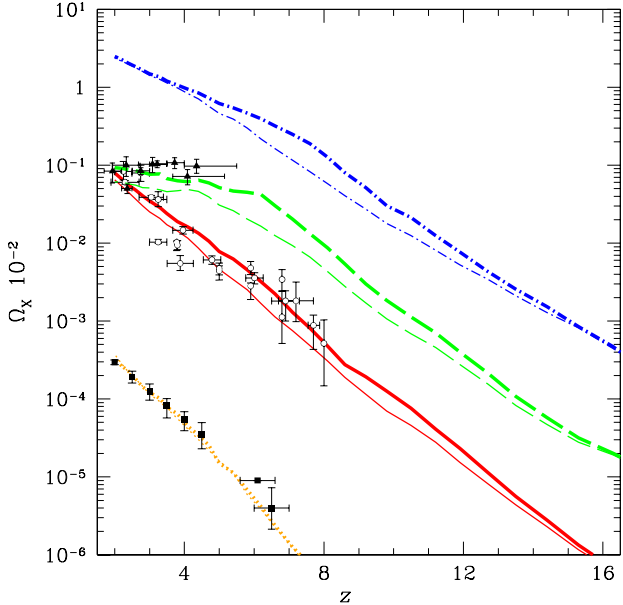,width=.55\textwidth,angle=0}}
\vspace{-0.5cm}
\caption{Mass density evolution of several phases X: MBHs
  (orange dotted lines; observational estimates in squares), \pIaII
  stars (red solid lines; empty circles), cold gas (green long-dashed
  lines; triangles), and hot gas in halos with normal galaxies (blue
  dot-dashed lines), for the same models as in Figure \ref{temp} (same
  line widths).}
\label{densities}
\end{figure}

\begin{figure}
\vspace{-0.7cm}
\centerline{\psfig{file=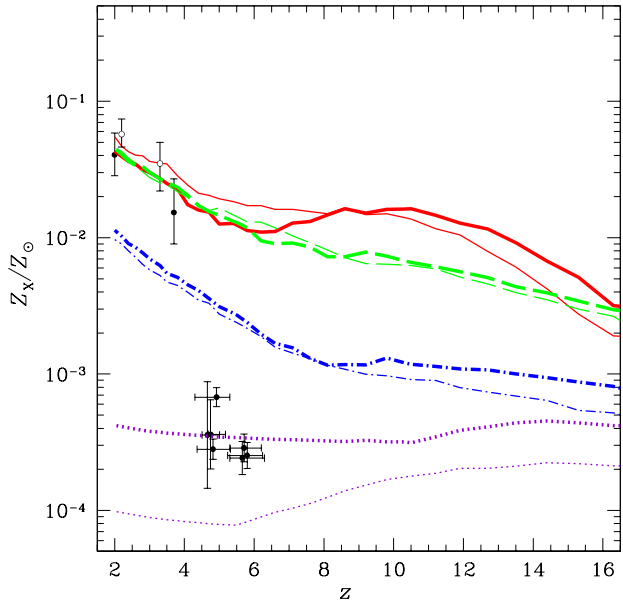,width=.55\textwidth,angle=0}}
\vspace{-0.5cm}
\caption{Average metallicity evolution in several phases X: ionized IGM
  (violet dotted lines; observational estimates in crosses), \pIaII
  stars (red solid lines; empty circles), cold gas (green long-dashed
  lines, triangles), and hot gas in halos with normal galaxies (blue
  dot-dashed lines), for the same models as in Figures \ref{densities}
  and \ref{metallicities} (same line widths).}
\vspace{0.2cm}
\label{metallicities}
\end{figure}

In Figures \ref{densities} and \ref{metallicities}, we show the
evolution of the main cosmic mass densities and mass-weighted
metallicities predicted by AMIGA in the same models as in Figure
\ref{temp}. We will comeback to these Figures in Section \ref{summ}.

\section{PHOTOMETRY}\label{out}

In the previous sections, we have described the modeling of the
temperatures, metallicities, and ionized fractions of the various IGM
phases, and the structural, chemical, kinematic, and dynamic
properties of luminous objects. This is enough for some applications
of the model (see e.g. \citealt{SM13}). However, for other applications,
the photometric properties of luminous objects must also be modeled.

\subsection{Galaxy Luminosities}\label{galprop}

In the case of normal metal-rich galaxies or low-mass \pIII stars,
AMIGA incorporates the evolutionary stellar population synthesis
models by \citeauthor{BC03} (\citeyear{BC93,BC03}, hereafter BC; see
also \citealt{CWB96}). These models use the isochrone synthesis
technique to compute the photometric properties of simple stellar
populations with a fixed IMF and metallicity. (See Sec.~\ref{SEDs} for
massive \pIII stars.) The SED of a stellar population of the same IMF
and metallicity, but arbitrary history of star formation, is computed
by means of a convolution integral of the spectrum of the simple
population and the desired star formation rate. The input for the BC
models is based on the evolutionary tracks of stars with different
masses and metallicities from \citeauthor{Bre94} (\citeyear{Bre94},
``the Padova tracks''), and the stellar spectra from the \citet{Kur79}
stellar model atmospheres. Energy fluxes are calculated for seven
values of the chemical composition: $Z=$ 0.005, 0.02, 0.2, 0.4, 1,
2.5, and 5 $Z_\odot$.

Using the BC models we can infer the luminosities, in different
photometric systems including wide and narrow bands in the observer
and galaxy frames, of galaxies with known stellar mass at formation
and star formation and chemical enrichment histories. The star
formation history of satellites is stored in the following discretized
form. We define a sequence of appropriate cosmic time steps inside
which the star formation rate is approximated by the corresponding
average value. The star formation history of a stellar population at
$t\obs$ is then simply given by the fraction $S\!F\c(j)$ of the final
mass $M\sfc$ of stars formed in each time step $j$. The final mass of
formed stars is the solution of the set of equations
(\ref{e1})--(\ref{e3}) at $t\obs$, while the fractions $S\!F\c(j)$ are
obtained by simple actualization, each time there is new star
formation or some old star mass is lost, of the galactic component
under consideration.

There is only the problem of memory limitations which puts severe
limitations on the total number of bins in the satellite array. This
is particularly annoying for the case of star formation histories
because the total number of distinct histories is a combinatorial
function of the number of time steps used. For this reason, we cannot
use a constant time step as narrow as desired (for instance, equal to
the typical duration of a starburst). However, taking advantage of the
fact that the luminosities at high frequencies of any stellar
population falls off soon after its formation while the luminosities
at low frequencies are little sensitive to the time elapsed since
then, we adopt only five time steps of appropriate varying width,
being the last two bins of the order of a starburst duration.

In principle, the chemical enrichment history of satellites can also
be discretized in the same way. Unfortunately, the total number of
different {\it joint} star formation and chemical enrichment histories
would be prohibitively large even for small numbers of metallicity
bins. For this reason, to calculate the luminosity of a stellar
population we assume a constant chemical enrichment rate, determined
by the final metallicity and total stellar mass, solutions of the set
of equations (\ref{e1})--(\ref{eIII}).

In this way, we can obtain the luminosity in any desired wide or
narrow band filter of any stellar population with one of the 7
specific metallicities for which the BC models are available or any
other metallicity by interpolation among them. We stress that the
approximations of discrete star formation histories and constant
chemical enrichment rate affect just the final photometry of stellar
populations of satellite galaxies, not even their stellar ages, which
are accurately calculated. They do not affect either the evolution of
their total stellar masses and total metallicity, which are accurately
monitored. We want also to mention that the version of the BC model we
use has been adapted to include the K-correction as well as the
correction for redshift-dependent absorption due to intervening
neutral hydrogen (the Ly$_{\alpha}$ forest is modeled according to
\citealt{Meik06}). It has also been adapted to provide the rest-frame,
extinction-free flux of ionizing photons and associated energy emitted
by galaxies, necessary for the calculation of the reionization and
associated reheating of the IGM. The extinction in narrow-band
photometry is also available for several important lines. The
obscuration by dust is modeled taking into account the usual
prescription that the optical depth is proportional to the metal
column density \citep{KTN10}, with wavelength-dependent
proportionality factors taken as free parameters to be adjusted once
for all by comparing the predictions of AMIGA with observations making
use of the same filters.

\subsection{AGN Luminosities}\label{AGN}

To compute the photometric properties of AGN we need not only the
masses of the associated MBHs and the rate at which they accrete
matter, mentioned in Section \ref{BHG}, but also a radiation model of
these objects. In AMIGA we adopt the simple model developed by
\citet{Hea03} assuming that the radiative pressure onto the infalling
gas, opposed to the gravitational pull by the MBH, produces damped
oscillations that quickly reach a stationary regime. The AGN
bolometric luminosity is then fully determined by the MBH feeding
rate, according to
\begin{equation}
L(t)=\frac{1}{2}\dot M\BH^{\rm g} V\last^2\,,
\label{BolL}
\end{equation}
where $V\last^2\approx c^2/K\last$ is the squared velocity at the last
marginally stable Keplerian orbit around the MBH,\footnote{This
  expression is only valid provided the Eddington efficiency is
  neglected with respect to one. In AMIGA we account for the exact
  expression leading to an AGN luminosity that is non-linear in $\dot
  M\BH^{\rm g}$ \citep{Hea03}.} with radius equal to $K\last=9.2$
times the Schwarzschild radius \citep{Hea03}. 

The bolometric luminosity function of AGN is computed as the product
of the time each single source spends in the wanted luminosity bin,
known from the luminosity curve (eq.~\ref{BolL}), times the specific
AGN reactivation rate \citep{Hea03}. In case of AGN activated in
galaxy mergers or in direct cooling flows such a reactivation rate
coincides with the formation rate of halos times the typical total
number of central galaxy mergers or disk-instability episodes taking
place after the halo forms. While in case of AGN activated in tidal
interactions among satellites, the distribution of bolometric AGN
luminosities is directly related to the distribution of times elapsed
since the corresponding (Poissonian distributed) satellite
interactions.

Once the bolometric luminosity of AGN is known, their spectrum
described in Section \ref{AGNf} readily leads to their luminosities in
any desired observer-frame photometric band, which is then properly
corrected for extinction by \HI according to \citet{Meik06}. AMIGA
assumes that the only period AGN are visible, after correction for
dust obscuration according to \citet{GWAW04}, is after the accretion
rate has reached its maximum value and the dispersion of gas is the
most marked. Before that moment, they are completely enshrouded within
the gas cloud.

\section{SUMMARY, FIRST RESULTS, AND DISCUSSION}\label{summ}

AMIGA is a very complete, detailed, analytic model of galaxy formation
devised to account fully consistently for the coupled evolution of
luminous objects (galaxies and AGN) and IGM since the dark ages. It
incorporates molecular cooling and \pIII stars, the luminous objects
with the most dramatic feedback, and accurately accounts for the
intertwined evolution of the halo MF and the IGM temperature and
ionization state.

AMIGA treats all those aspects of galaxy formation that can be
causally linked to each other and to the underlying cosmology (DM
clustering, halo and hot gas structure and kinematics, cooling, disk
formation, and BH growth) without any free parameter. The only free
parameters in the model, \\ Hot gas and IGM:\\ \indent $b_{\rm cl}$:
\HII clumping bias\\ \indent $h\rec$: hot gas recycling
fraction\\ \pIII stars: \\ \indent $p\III$: yield of massive
stars\\ \indent $\beta\III$: stellar mass fraction ending locked in
BHs\\ Normal galaxies:\\ \indent $\alpha\G$: star formation efficiency
\\ \indent $f\esc$: escape fraction of ionizing photons\\ \indent
$\rho\dis$: critical dissipation density\\ Galaxy interactions:
\\ \indent $\deli$: minimum relative mass for interactions
(0.01)\\ \indent $\chi\DB$: disk-to-bulge mass transfer efficiency
(0.01) \\ AGN:\\ \indent $\epsilon\AGN$: quasar-mode heating
efficiency\\ concern poorly known aspects (small-scale ionized gas
distribution, stellar and AGN feedback, spheroid structure, and galaxy
interactions) that are disconnected from each other.

Contrarily to the usual procedure, no free parameters are used to
specify the initial conditions (IGM metallicities, temperatures, and
ionization state, MBH masses, ionizing UV fluxes,...). The modeling
starts from trivial initial conditions at the dark ages, and follows
the formation of the first generation galaxies with \pIII stars,
characterized by only two parameters fixing at the same time their IMF
and feedback ($p\III$ and $\beta\III$ or, alternatively, $f_2$ and
$f_3$; eqs.~[\ref{mu}]--[\ref{y3}]).

The fact that there is neither any artificial freedom nor poorly
motivated initial conditions that may spuriously facilitate the
fitting of observations renders the predictions of AMIGA particularly
reliable.

But the number of free parameters is yet quite large. A first
analysis of the parameter space reveals that, if the parameters
characterizing \pIII stars take values outside some `acceptability
ranges', the IGM metallicity never becomes high enough to trigger the
formation of normal galaxies and MBHs. The acceptability range of each
of these parameters is very robust in the sense that it is independent
of the value of its partner.

From the meaning of the different parameters and the way \pIII stars
form, it is clear that the properties of these stars are fully
determined by the parameters in the corresponding set. No other
parameter can influence them. However, the situation is different for
the properties of normal galaxies and MBHs. They depend of course on
the parameters in their respective sets, but they may depend on the
properties of \pIII stars as well. Indeed, \pIII stars are responsible
for the metal enrichment of the IGM that is incorporated into
halos. As the higher the metallicity of the hot gas, the higher the
cooling rate, the properties of \pIII may influence not only galactic
metallicities, but also the amount of cold gas falling into galaxies
and, hence, the structural properties of these objects. On the other
hand, \pIII stars reheat the IGM, increasing its entropy floor and,
hence, the minimum mass of halos able to trap gas. As the larger the
mass of a halo, the smaller its concentration, and the less intense is
cooling, the amount of cold gas feeding the most abundant dwarf
galaxies will be smaller. The question then rises: do the properties
of \pIII stars influence those of normal galaxies?

To answer this fundamental question we compare next the results of
AMIGA drawn from the two plausible models differing only in the values
of the \pIII star parameters, used in the previous Figures. All the
remaining parameters are taken with identical values,\footnote{We have
  the right to chose them so despite the different values of the \pIII
  star parameters because, as mentioned, all {\it parameters} in the
  previous list are disconnected from each other.} so any possible
difference in the final properties of normal galaxies will be due to
the influence of \pIII star properties.

As shown in Figure \ref{metallicities}, the higher metallicity of the
ionized IGM found in the case of the top-heaviest \pIII star IMF (with
the largest $f_2$ value) affects the metallicities of the hot gas,
cold gas, and stars, which are a little higher in this model than in
the one with less top-heavy \pIII star IMF (with the smallest $f_2$
value and, hence, lower IGM metallicity). But normal galaxies,
particularly dwarf ones, eject large amounts of metals in the hot gas,
so its metallicity in any given halo quickly increases. Nonetheless,
halos keep on accreting low-metallicity IGM and there are always new
halos accreting gas for the first time, so the average hot gas
metallicity increases very slowly, causing a similar trend to the
metallicities of stars and cold gas in galaxies. Due to the
permanently renewed memory of the IGM metallicity, the convergence of
the metallicities in the two models is delayed until $z\sim 8$ (a
little later in the case of stars due to their larger memory on their
past history; the cold gas mass is instead continuously renewed). 

The {\it structural} properties of normal bright galaxies and MBHs, as
traced by the masses of their galactic components and phases, show a
similar behavior. As shown in Figure \ref{densities}, they are even
less sensitive to the properties of \pIII stars at very high
$z$'s. The reason for this is that atomic cooling is much more
sensitive to the density of the hot gas than to its metallicity, while
the hot gas density is very similar in the two models because of the
very similar temperature of the corresponding ionized IGM (see
Fig.~\ref{temp}). Indeed, in the model with the top-heaviest \pIII
star IMF, ionized bubbles grow more rapidly owing to the larger
abundance of massive \pIII stars and, hence, the larger rate of
ionizing photons. But the temperature reached by IGM in bubbles is
essentially the same in the two models because of the mass-independent
SED of high-mass \pIII stars. 

The structural properties of normal bright galaxies in the two models
are so similar that, despite the different masses of MBH seeds in the
two models (through $f_3$), no significant difference is found in the
MBH mass densities (see Fig.~\ref{densities}). The reason for this is
that, although the masses of the coalesced \pIII BH remnants depend on
the \pIII star IMF, as soon as spheroids begin to grow, MBHs accrete
such large amounts of gas compared to the mass of their seeds that
MBHs rapidly lose the memory of those seeds.

Therefore, one fundamental result of AMIGA is that the structural
properties of normal bright galaxies and MBHs essentially decouple
from those of \pIII stars; there is only a small coupling in the
metallicities. In this sense, SAMs and simulations with
non-self-consistent initial conditions should correctly predict the
properties of normal bright galaxies and MBHs in the low and
moderately high $z$ Universe or even in the high-$z$ one provided we
do not care about metallicities. This justifies previous studies on
galaxy formation focusing on the properties of nearby galaxies
achieved by means of models with inaccurate initial conditions. The
situation is completely different, however, if one is interested in
predicting accurate galactic metallicities or accurate properties, at
any $z$, of small luminous objects (\pIII star clusters and normal
dwarf galaxies), or still if one is interested in the evolution of the
Universe at very high-$z$ where the effects of \pIII stars are the
most marked. Then, the use of a model like AMIGA is mandatory.

\begin{figure}
\vspace{-0.7cm}
\centerline{\psfig{file=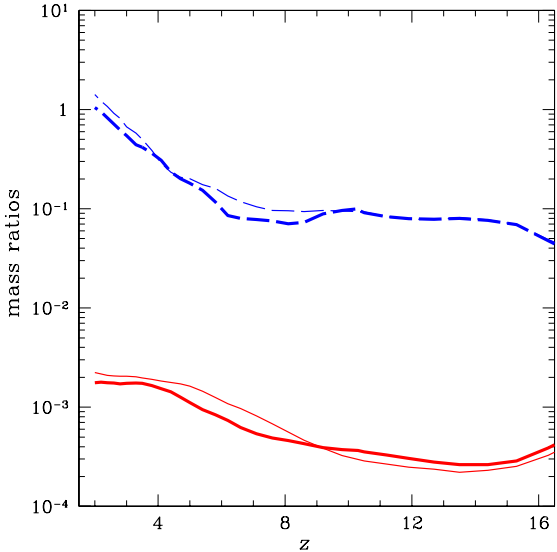,width=.5\textwidth,angle=0}}
\vspace{-0.5cm}
\caption{Evolution of the MBH to spheroid (red solid lines) and
  spheroid to disk (blue dashed lines) global mass ratios for the same
  models as in the preceding Figures (same line widths).}
\vspace{0.2cm}
\label{ratios}
\end{figure}

The results of the two models analyzed also show that spheroids grow
in parallel to MBHs, and disks grow in parallel to spheroids, so that
the MBH to spheroid and spheroid to disk mass ratios are kept rather
constant (see Fig. \ref{ratios}). The former effect is the consequence
of star formation in spheroids being quenched by the ISM reheating by
AGN, whose bolometric luminosities are self-regulated by the MBH
feeding. The latter is the consequence of the fraction of cold gas
going into disks or spheroids, which depends on the spheroid mass.

The constant MBH to spheroid mass ratio is first reached in massive
galaxies because, when dwarf galaxies form, MBHs are tiny. This is the
reason why the average MBH to spheroid mass ratio shows a small
increasing trend with decreasing $z$. Of course, for the stationary
regime in both ratios to be reached, all galactic components must
develop freely. At low-$z$'s ($\la 7$), cooling becomes increasingly
inefficient and disks begin to ted, while spheroids keep on growing
through galaxy mergers that become increasingly dry. The spheroid to
disk mass ratio then begins to increase with decreasing $z$. However,
the MBH to spheroid mass ratio is kept unaltered. The reason for this
is that, in dry mergers, both MBHs and spheroids continue to grow, the
former through the coalescence of the progenitor MBHs, and the latter
through the addition of the stellar populations of the merging
galaxies. As the MBH to (stellar) spheroid mass ratio of the two
progenitors is the same, so is also the ratio in the final
object. This explains why the MBH to spheroid mass ratio is kept
unchanged until $z=0$.

Therefore, a second fundamental result of AMIGA is that the growth of
all galactic components is ultimately controlled by that of MBHs. In
fact, these objects play a crucial role in the evolution not only of
normal galaxies, but also, through the feedback of those objects, in
the evolution of all cosmic properties. For this reason, any model or
simulation of galaxy formation must necessarily deal self-consistently
with MBHs. This is true even regardless of whether or not the model or
simulation includes accurate MBH seeds.

The total number of free parameters in the list above is 10. However,
some of them can be removed if one concentrates in the high-$z$
Universe (at, say, $z\ge 2$). Galaxy interactions play indeed a
significant role only in the detailed morphological appearance of
galaxies. As the galactic morphologies are unresolved at very
high-$z$, one can study in a first step the formation of the first
luminous objects, taking those two parameters with fixed reasonable
values (like those quoted in parentheses in the list above), and
adjust them in a second step by studying the local universe.

The total number of free parameters in the first step then reduces to
8. Although this number is still considerably large, it is smaller
than the number of independent data sets available on the cosmic
properties at $z\ge 2$ (see \citealt{SM13}), so the problem is
well-constrained. Moreover, the fact that the structural properties of
normal bright galaxies, the only observable at those $z$'s, are
independent of the properties of \pIII stars translates into the
decoupling of their respective parameters. As shown in \citet{SM13},
this notably simplifies the adjustment of the free parameters of the
model. In fact, a very simple fitting procedure can then be devised,
where all the parameters are adjusted sequentially through the fit of
one independent data set each. This has the added advantage of
rendering the complex process of galaxy formation particularly easy to
comprehend.

\begin{acknowledgments}
This work was supported by the Spanish DGES grant
AYA2012-39168-C03-02, and the Catalan DIUE grant 2009SGR00217.  GB
acknowledges support for this work from the National Autonomous
University of M\'exico, through grants IA102311 and
IB102212-RR182212. We are thankful to Guy Mathez for his inestimable
encouragement to start this project.
\end{acknowledgments}

\begin{appendix}

\section{Dissipative contraction}

Dissipative contraction of spheroids during star formation is due to
the loss of orbital energy by the dense nodes of cold gas where stars
form, hereafter the gas clouds, that move inside the spheroid. Gas
clouds also radiate {\it internal} energy as they contract and
fragment to form stars, but this energy loss does not alter the cloud
orbits. For simplicity, we neglect the adiabatic contraction (or
expansion, in the final gas ejection) of the dissipationless component
(stars and DM).

The specific kinetic energy of the gas associated with the orbital
velocity of clouds supporting the spheroid is
\beq
e=-2f\frac{GM\B}{r\B}=-\frac{8\pi}{3} g G \bar \rho\B r\B^2\,,
\eeq
where $\bar \rho\B$ is the mean spheroid density and $f$ and $g$ are
two constants dependent on the specific spheroid density profile. The
emission power per unit gas mass at the base of the dissipative
contraction of the system can be assumed to satisfy the simple
equation
\beq 
\frac{\der e}{\der t}=-\epsilon\dis\frac{3kT\cgb}{2\mu m\p\tau\dis}\,,
\label{dissip}
\eeq 
where $\epsilon\dis$ is the dissipation efficiency, $T\cgb$ is the
effective temperature of the gas accounting for the orbital kinetic
energy of clouds, and $\tau\dis$ is the dissipation timescale.

To figure out the expression for $\tau\dis$ we can take into account
that the orbital energy radiated {\it per unit gas mass} is also equal to
\beq 
\frac{\der e}{\der t}=- f\dis \Lambda[T\cgb(t),Z\cgb(t)]
\frac{\bar n\cgb(t)}{\mu m\p}
\label{dissip2}
\eeq 
where $\bar n\cgb$ is the mean particle density in the spheroid, and
$f\dis$ is the fraction of the radiated energy that can be associated
with the orbital motion of clouds. The cooling function
$\Lambda(T\cgb,Z\cgb)$ for a gas at $T\cgb$ of order $10^5$ K, as
corresponding to halos with the relevant masses, and $Z\cgb$ spanning
from $10^{-2}$ Z$_\odot$ to 1 Z$_\odot$ is proportional to $T\cgb$ with
proportionality factor inversely proportional to the square root of
$Z\cgb$ \citep{SD93}, that is, $\Lambda(T\cgb,Z\cgb)\approx
3kT\cgb/[2n\ch\tau\ch(Z\cgb)]$, where $\tau\ch(Z\cgb)\approx
\tau\ch(Z_\odot)(Z\cgb/Z_\odot)^{-1/2}$ for suited values of the
characteristic number density $n\ch$ and time $\tau\ch(Z_\odot)$. We
thus have $\tau\dis\approx
(\epsilon\dis/f\dis)\tau\ch(Z\cgb)n\ch/\bar n\cgb$.

Substituting this expressions for $\tau\dis$ into equation
(\ref{dissip}), taking into account the virial relation $3kT\cgb/(\mu
m\p)=fGM\B/r\B=(4\pi/3) g G \bar \rho\B r\B^2$, we arrive at the
following approximate equation for the scale radius at a time t after
the beginning of star formation and dissipation,
\beq 
r\B^2\,\der r\B= -\frac{3\epsilon\dis\,f}{16\pi f\dis\,g} \left[\frac{Z\cgb(t)}{Z_\odot}\right]^{1/2}\frac{M\cgb(t)}{n\ch}\frac{\der t}{\tau\ch(Z_\odot)}= 
-\frac{Z\cgb^{1/2}(t)M\cgb(t)} {Z_\odot^{1/2}\rho\dis}\frac{\der t}{\tilde\tau\acc},
\label{eq1}
\eeq 
where $\tilde\tau\acc$ is the universal time interval elapsed between
the formation of the spheroid and the moment when star formation is
quenched due to the reheating and ejection of the remaining gas by
AGN, and parameter $\rho\dis$ is a critical dissipation density where
all (known and unknown) constant factors are encapsulated.  Equations
(\ref{eq1}), (\ref{e2}), and (\ref{eII}) for C=B determine the coupled
evolution of the contracting scale radius and the cold gas mass and
metallicity in the spheroid.

\end{appendix}

\end{document}